# Entropy stabilized form chirality in curved rod nematics: structure and symmetries.

Alexandros G. Vanakaras, [a] Edward T. Samulski *[b] and Demetri J. Photinos [a]

[a.] *Department of Materials Science, University of Patras, 26504 Patras, Greece.*
[b.] *Department of Chemistry and Applied Physical Sciences, University of North Carolina, Chapel Hill, NC 27599-3201 USA.*
[c.] ** Corresponding author; Email: et@unc.edu*

Monte Carlo molecular simulations of curve-shaped rods show the propensity of such shapes to polymorphism revealing both smectic and polar nematic phases. The nematic exhibits a nanoscale modulated local structure characterized by a unique, polar, $C_2$-symmetry axis that tightly spirals generating a mirror-symmetry-breaking orgnization of the achiral rods—form chirality. A comprehensive characterization of the polarity and its symmetries in the nematic phase confirms that the nanoscale modulation is distinct from the elastic deformations of a uniaxial nematic director in the twist-bend nematic phase. Instead it is shown that, analogous to the isotropic-to-nematic transition, entropy stabilizes the roto-translating polar director in the polar-twisted nematic phase. The conflation of macroscale form chirality in ferroelectric nematics with that in the twist-bend nematic stems from the misattribution of the nanoscale modulation in the lower temperature nematic "$N_X$ phase" found in CB7CB dimers.

## 1. Introduction.

Most nematic-forming molecules (nematogens) self-organize into the uniaxial, apolar and achiral, nematic phase ($N_U$). Nematic polymorphism—structural transformations within the simplest liquid crystal (LC)—was foreseen by Born in 1916 when he proposed that fluid ferro- and antiferro-electric structural polymorphs can account for LC formation (*1*). Born's model was inspiring but ultimately unsound as electrostatic interactions are overwhelmed by thermal energy. Nevertheless, reinforced by the work of A. Khachaturyan (*2*), the concept of a polar, ferroelectric nematic phase ($N_F$) remained an elusive "holy grail" until recently, when independent reports of two nematic phases using achiral calamitic (rod-like) nematogens appeared (*3*, *4*). The lower temperature phases are $N_F$ polymorphs (*3*, *5*–*7*) having "form chirality"—a twisting supramolecular structure with broken mirror symmetry (Fig. 1a) (*8*, *9*). This constitutes a substantial advance, since the only ferroelectric LCs known previously were chiral molecules that form smectic SmC* phases (Fig. 1b) (*10*). We revisit Monte Carlo simulations of achiral, curve-shaped rods using the identical parameterization of earlier researchers, and we show incontrovertible evidence confirming that the structure of the modulated nematic phase is that of the theoretically predicted (*11*, *12*) polar-twisted nematic polymorph ($N_{PT}$), exhibiting a mirror symmetry-breaking helicoidal supramolecular structure with nanoscale modulation. Moreover, that structure is derived solely from excluded volume interactions independent of continuum-elastic considerations; at higher densities, an antiferroelctric smectic structure forms.

### 1.1 Entropic stabilization.

Seventy-five years ago Onsager showed that entropy drives the global I-$N_U$ (disordered isotropic to ordered nematic) phase transition using hard, rod-like particles with aspect ratio *L/D* (*13*); the translational entropy increase in $N_U$ more than compensates the loss of orientational entropy in the I phase (*14*). In his athermal model, I-$N_U$ occurs above a critical, *L/D*-dependent, density explaining experimental observations in TMV (*15*) and polypeptide solutions (*16*). At higher densities the $N_U$ transforms into a stratified smectic phase (Sm A)(*17*).
Recently, a renaissance in the study of the LC phases in mineral colloidal dispersions (*18*) has occurred, showing that rodlike particles form a $N_U$ and board-like geometries form biaxial nematic and smectic phases (*19*, *20*). Particle curvature impacts both local and long-range organization (*21*–*24*) with bent-core mesogen shapes yielding impressive demonstrations of smectic polymorphism (*25*). Also, it has been established that the polar shape of bent-core mesogens promotes long-range polar molecular organization and a ferroelectric response (*26*, *27*).

### 1.2 Mirror symmetry breaking

Robert Meyer (*28*) was the first to conjecture that achiral nematogens could spontaneously adopt a chiral, mirror symmetry-breaking structure wherein the $N_U$ nematic director $\hat{n}$ twists to relieve flexoelectric polarity generated by bending

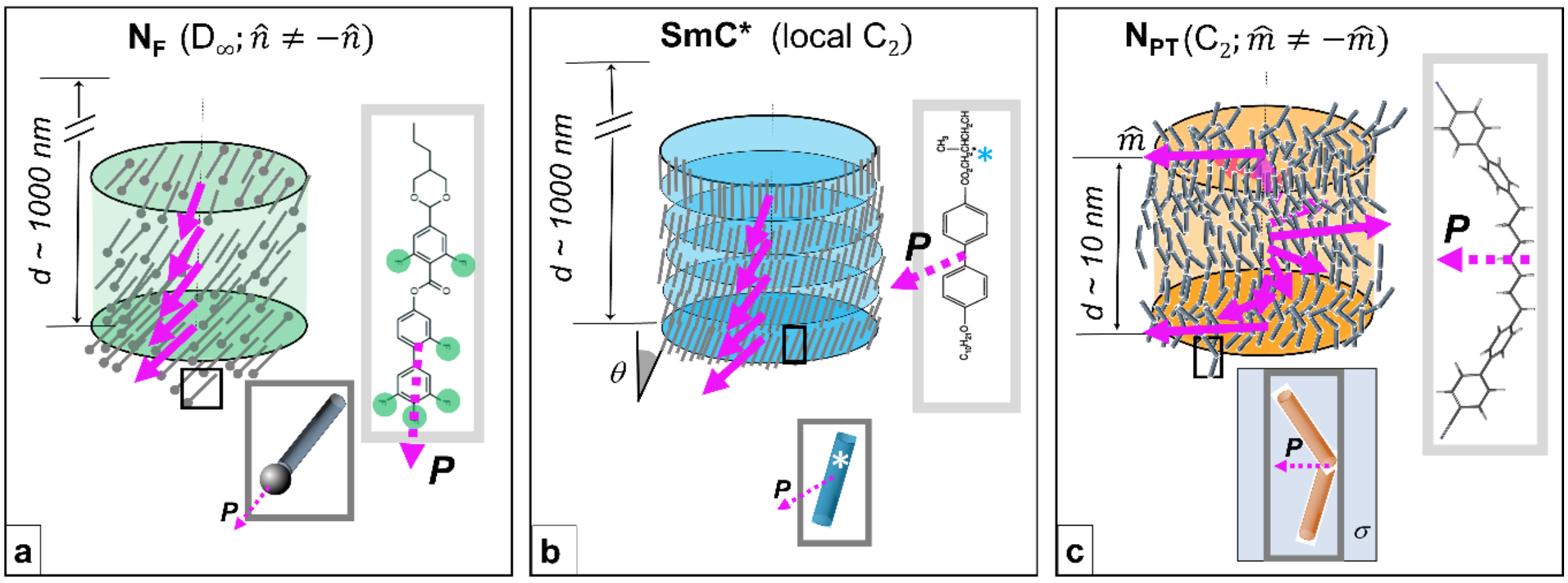

**Figure 1 Typical liquid crystal phases showing local polar ordering of the director $\hat{n}$ combined with chirality. a.** Ferroelectric nematic phase ($N_F$); **b.** Chiral smectic C phase (SmC*); **c.** Polar twisted nematic phase ($N_{PT}$). Representative molecular structures are shown for each case with primitive molecular models showing the molecular polarity vectors *P*. The nematic phases in (a) and (c) consist of achiral molecules and show form chirality. The smectic phase in (b) consists of chiral molecules, self-organizing into intrinsically chiral ensembles. In all cases the local polarity remains perpendicular to the modulation (roto-translation) direction. The length scale of the modulation in (c) is two orders of magnitude smaller than that in (a) and (b). Unwinding the "soft" (large pitch) modulation in the SmC* generates a ferroelectric state. No such unwinding has ever been reported for the nano-modulated $N_{PT}$ phase.

$\hat{n}$ ; Meyer called that macroscale elastic distortion the *twist bend nematic*, $N_{TB}$ .

Re-examination of polar-shaped odd homologues of symmetric cyanobiphenyl dimers (Fig. 1c) has revealed two nematic phases (*29*). In 1991 an apparently conventional $N_U$ in CB9CB was observed to transform to a second nematic phase on cooling (*30*). Subsequently that phase was termed $N_X$; in other odd-n homologues the $N_X$ was shown to have a helicoidal structure with a nanometer-scale pitch (Fig. 1c) (*31*, *32*). That tightly wound form chirality has a superficial resemblance to Meyer's $N_{TB}$, a loosely spiralling, elasticity-driven, micron-scale, spontaneous deformation of $\hat{n}$ in the $N_U$(*28*). That similarity, as pointed out and discussed in detail elsewere(*33*), has led to an erroneous identification of the $N_X$ phase with Meyer's $N_{TB}$, and a flawed rationalization: namely, that all helical modulations in nematics are elastically driven, see refs (*33*–*35*).

### 1.3 Motivation and analyses

Establishing the molecular prerequisites for polar nematic polymorphism and characterizing polar organization across different length-scales remains a challenge. To address this issue and to obtain valuable insights into molecular features motivating nematic polymorphism, we performed simulations of sterically-interacting, achiral, curved rods using model parameterization identical to that used in prior simulations(*36*). We critically assess and analyse the simulation results with an array of new order parameters and spatial correlation functions to reveal the underpinnings of the nanoscale helical pitch (form chirality). The results support the structure of a new nematic polymorph, the polar-twisted nematic ($N_{PT}$) first advanced in 2016 (*11*).

Our new findings are based on an extant model of curved rods introduced by C. Greco and A. Ferrarini (*37*) and subsequently studied by Chiappini and Dijktsra(*36*). Using the same parameterization we critique the interpretation of the latter researchers of the nematic phase; we also show that their results at higher densities are flawed because of improper sizes of the simulation boxes.

We show that the nano-modulation in this new nematic polymorph is not consistent with the Frank-Oseen continuum-elastic interpretation of spontaneously formed chirality of the type responsible for the $N_{TB}$. Rather, simulations (with correct box sizes) confirm both the experimental attributes and the predictions of the $N_{PT}$ model for the second low-temperature nematic phase observed in bent-core mesogens.

In order to confirm the predictions of the $N_{PT}$ model, we introduce a complete and consistent framework for the analysis of the local and global structure as well as the symmetries of the nano-modulated structure. We obtain quantitative results for the full set of order parameters describing the lower temperature nematic polymorph. We discuss why the continuum elasticity description of nano-modulated nematics prevents a correct analysis of the simulation results. Additionally, we quantitatively calculate the pressure (density) dependence of: i) the local biaxiality of the simulated nano-modulated nematic phase; ii) the different "tilt" angles associated with the principal axis frames of the 2nd rank tensorial molecular properties; and iii) the vector-pseudovector order parameter which constitutes a direct measure of the form charility of the system. Finally, we reproduce, based on the 2016 theoretical model of the $N_{PT}$, nearly exactly all the pair correlation functions along the axis of modulation in terms of the order parameters of the system.

### 1.4 Summary of findings

Our simulations show that rod curvature results in a multiplicity of new phases including the $N_{PT}$ as well as antiferroelectric smectic phases. Significantly, we find form chirality in $N_{PT}$ to be an entropy-stabilized, modulated nematic polymorph that has a unique polar $C_2$-symmetry axis $\hat{m}$ normal to the modulation axis; i.e., $\hat{m}$ is a periodic, roto-translating, polar director.

New order parameters and correlation functions quantify modulations of local polarity. We present descriptions that illuminate the physics underlying form chirality in nematics having different local symmetries, i.e., $N_F$ ($D_\infty$; $\hat{n} \neq - \hat{n}$ ), $N_{TB}$ ($D_{\infty h}$;

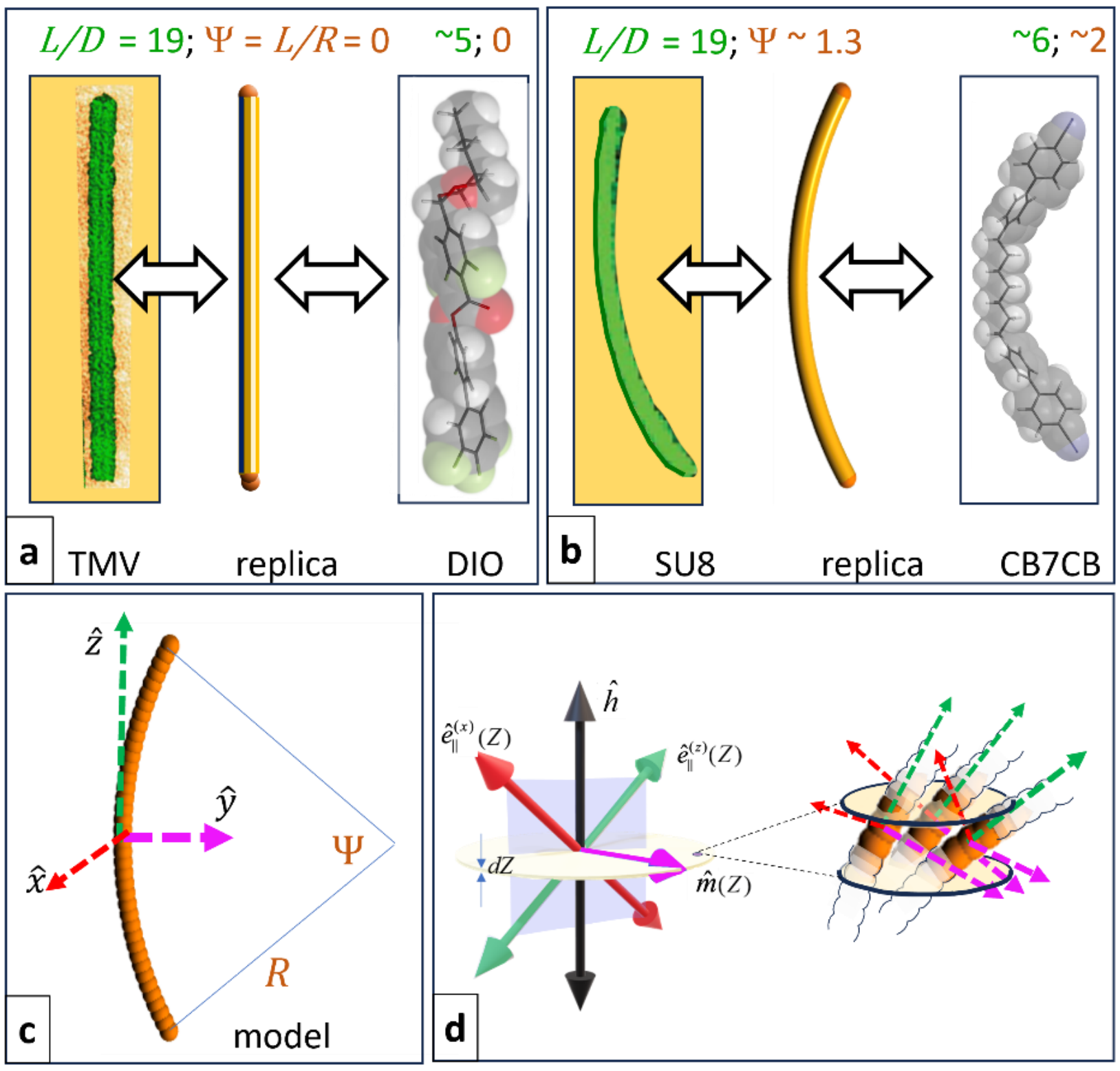

**Figure 2 Colloidal, thermotropic and model mesogens. a.** Onsager rigid rod limit, e.g., TMV (Tobbaco Mosic Virus), sphero-cylinder $L/D=19$, DIO ($N_F$ liquid crystal) $L/D\approx 5$; **b** SU8 curved rod photoresist [20] and its replica ( $L/D=19$ and sector angle $\Psi\equiv L/R=1.31$ ); CB7CB dimer LC. The curved rods are treated as a collection of 39 spherical beads of diameter *D* and capped with two half-spheres. **c** the particle/molecule frame has $\hat{y}$ along the $C_2$ symmetry axis of the model and $\hat{x},\hat{z}$ normal to the two symmetry planes; **d** Structure and symmetry in the modulated nematic with the directions of maximum alignment ( $\hat{e}_{\parallel}^{(x)}$, $\hat{m}\equiv\hat{e}_{\parallel}^{(y)}$, $\hat{e}_{\parallel}^{(z)}$ ) of the molecular axes ( $\hat{x},\hat{y},\hat{z}$ ) in a slab, perpendicular to the axis of modulation $\hat{h}$, of thickness $dZ$; the inset (right) shows a fragment of the local structure.

$\hat{n}$ = - $\hat{n}$ ), and $N_{PT}$ ( $C_2$; $\hat{m}\neq-\hat{m}$ ) phases, thereby facilitating mesogen design for new applications. Identical simulation parameterization contradicts previous claims (*36*) that Frank-Oseen nematic elasticity underlies nanoscale helical modulation, and conclusively shows that some of the earlier reported smectic-like phases are artefacts due to improper simulation-box dimensions.

In the following sections we describe the curved rod model and give the simulation details. We introduce a set of order parameters and correlation functions that are sensitive to form chirality (Section 2.1). Then, we critically assess the simulation results, e.g., we delineate the phase boundaries (Section 3.1) and use the new order parameters and correlation functions to quantify the associated structure in the observed nematic phase (Section 3.3). In Section 4 we discuss the regimes of validity for earlier simulations and critique the attribution of the underlying physics of the nanoscale modulations to continuum elasticity. In summary, we find that the polar-twisted nematic accounts for all the observations connected with the structure as well as the ordering and the symmetries in the lower temperature nematic phases found in odd-n linked CBnCB dimers. Those include the molecular scale helical modulation, the characteristic doubling observed in the NMR spectra of deuteriated dimers or solutes dissolved in the $N_{PT}$, and the possibility of a direct I-to-$N_{PT}$ phase transition. There are also specific implications for the $N_{PT}$ polymorph on the physical properties of these dimer mesogens, such as their dielectric and optical responses, their surface anchoring motifs, and their elastic deformations (on the appropriate, micron scale) that are distinctively different from those associated with an elastic twist-bend interpretation(*38*).

## 2. Modelling and simulations.

Molecular simulations of bent-core mesogens and colloidal particles employ a variety of models ranging from detailed atomistic models (*12*, *39*, *40*) to primitive models based on idealized, sterically interacting, geometrical particles (*36*, *37*, *41–43*). A primitive representation of the latter is a curved rod with diameter *D* and length $L$ forming an arc of radius $R$. The model parameters are $L/D$ and the sector aperture angle $\Psi\equiv L/R$ (see Fig. 2c and Supplementary Information SI.1). The model provides accurate descriptions of colloidal particles whose suspensions exhibit LC behaviour (*21*, *22*, *44*).

The curved-rod is modelled in our simulations as a collection of fused hard spheres of diameter $D$ with their centers evenly distributed on the arc of length $L$ of a circle of radius $R$. For the simulated systems with $L/D=19$ the rods are formed by 39 fused hard spheres of diameter $D$. Their molecular volume $v_m=\frac{\pi D^3}{6}\left(1+\frac{3}{2}\frac{L}{D}\right)$ is taken equal the volume of the corresponding smoothly curved spherocylinder shown in Fig. 2(b).

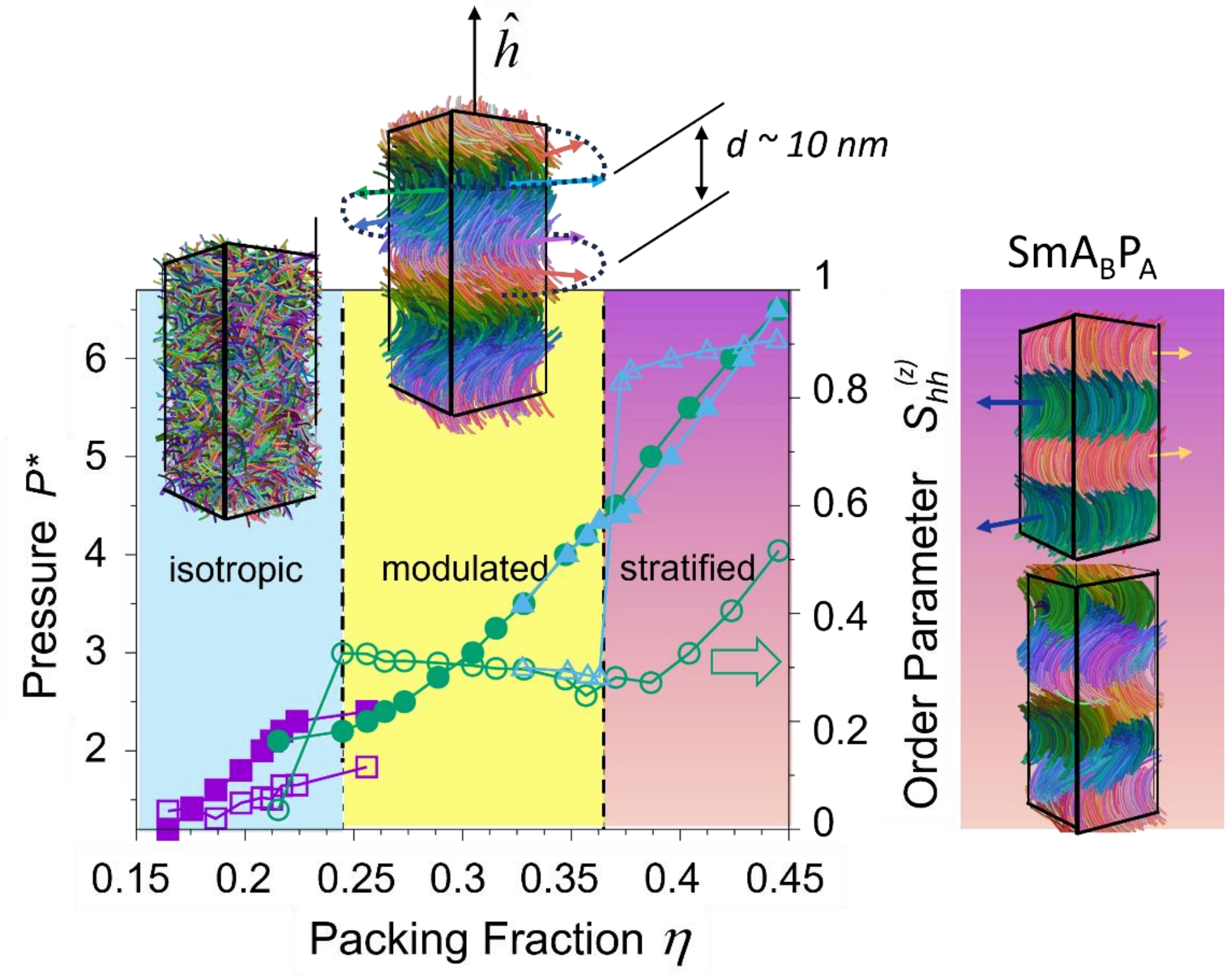

**Figure 3. Phase diagram of curved hard rods.** The aperture angle is $\Psi = 1.31$ . Solid symbols are pressure *vs* packing fraction equation of state; squares correspond to a compression sequence from the isotropic state. Circles correspond to expansion and subsequent compression from a nematic state at $P^* = 3$ . Triangles correspond to expansion from a close packed, highly ordered state at $P^* = 6.5$ . Open symbols refer to the order parameter $S_{hh}^{(z)}$ of the molecular *z*-axis obtained from expansion/compression runs (solid symbols). The vertical dashed lines indicate the low density bounds of the modulated nematic with periodicity *d* and the stratified smectic phases; the top right $SmA_BP_A$ phase is an orthogonal biaxial smectic phase ($SmA_B$) consisting of polar layers with an antiferroelectric arrangement ($P_A$); on compression the stratified phase appears in the form of self-organized ordered fragments (lower).

The spontaneous mirror-symmetry breaking in bent-core nematics, and nanoscale helical organization have been demonstrated by the purely-repulsive, curved-rod model, initially by Greco and Ferrarini (*37*) assuming soft repulsions, and later by Chiappini and Dijkstra for strictly hard-body systems (*36*). The former (*37*) simply validated a theoretical model (presented in the same paper and introducing explicitly local polar order) for the $N_X$ phase. The latter (*36*) explored structure in a nano-modulated nematic phase and its underlying driving forces; the authors also considered LC phases that the $N_X$ phase can transform into on cooling/compressing. They claim that nano-modulation is elastically driven with structure and symmetries of a miniature (nanoscale) $N_{TB}$. We re-examine those claims with simulations using the same curved-rod parameterization and advance new analysis tools for characterizing structure and symmetries.

Systems of curved rods (Fig. 2c) are simulated via standard *NPT* Monte Carlo ( $N = 4048$ particles), in orthogonal boxes applying periodic boundary conditions. The reduced pressure $P^* = Pv_m / k_B T$ (where $v_m$ is the molecular volume and $k_B$ is Boltzmann's constant) was varied from 1 to 7, corresponding to equilibrium packing fractions, $\eta = 0.16$ to $0.45$. Our main results are focused on the $L/D = 19$ , $\Psi = 1.31$ system studied in (*36*) enabling a direct comparison. Further details are given in the section S.1 of the SI.

### 2.1 Quantification of molecular ordering.

The analysis of the simulation results is focused on the structure and symmetries of the modulated nematic phase. The ordering of the molecules is quantified in terms of low rank orientational order parameters. The efficient and unambiguous evaluation of these parameters directly from the simulation data is facilitated by using appropriate correlation functions. These are introduced below. Their direct evaluation from the simulations is presented in the next section.

The curved-rod particles/"molecules" in Fig. 2.c have $C_{2V}$ symmetry. Identifying $\hat{y}$ with the $C_2$ molecular axis implies invariance of all the physical properties of the ensemble with respect to the separate molecular axis inversions $\hat{x} \Leftrightarrow -\hat{x}, \hat{z} \Leftrightarrow -\hat{z}$ . Furthermore, for nematics with local $C_2$ phase-symmetry, invariance is implied with respect to the simultaneous axis inversion $\hat{h}, \hat{l} \Leftrightarrow -\hat{h}, -\hat{l}$ in a local phase-fame formed by the three mutually orthogonal axes $\hat{h}, \hat{l}, \hat{m}$ , in which $\hat{m}$ is identified with the two-fold local symmetry axis. Accordingly, the orientational ordering is described, in ascending tensorial rank, by the following non-vanishing ensemble averages, defining the respective sets of order parameters:

***Order parameters of first rank***. As a result of the combination of the local $C_2$ symmetry of the phase and the $C_{2V}$ symmetry of the bent-rod molecules, there is just one non-vanishing polar order parameter, $p_\perp \equiv \langle \hat{y}\cdot\hat{m}\rangle$.

***Order parameters of second rank.*** The second rank parameters describing the orientational order of the three molecular axes $\hat{a} = \hat{x},\hat{y},\hat{z}$ are generally

$$S_{AB}^{(a)} = \left\langle 3\left(\hat{a}\cdot\hat{A}\right)\left(\hat{a}\cdot\hat{B}\right) - \delta_{A,B}\right\rangle / 2\,, \qquad (1)$$

forming three symmetric and traceless ordering matrices. When expressed in the frame $\hat{h},\hat{l},\hat{m}$, only $S_{hh}^{(a)}, S_{ll}^{(a)}, S_{mm}^{(a)}, S_{hl}^{(a)}$ survive the implications of the local phase symmetry. Of these sets, only five elements are independent (See S.2 for details).

Each of ordering matrices, $S_{AB}^{(a)}$, can readily be diagonalized to obtain the principal axes and corresponding principal values: $\hat{m}$, being the local phase-symmetry axis, automatically constitutes a principal axis of all the second rank tensors; therefore the other two principal axes, *e.g.*, $\hat{h}_p^{(a)}, \hat{l}_p^{(a)}$ are obtained by a single rotation of $\hat{h},\hat{l}$ about $\hat{m}$ by an angle $\theta^{(a)}$ satisfying the condition $S_{h_p^{(a)} l_p^{(a)}}^{(a)} = 0$. Obviously, the values of $S_{mm}^{(a)}$ remain unaffected by this rotation. The three principal values $S_{h_p h_p}^{(a)}, S_{l_p l_p}^{(a)}, S_{mm}^{(a)}$ are not independent as their sum vanishes. Accordingly, the ordering tensor in its principal axes frame (see fig. 2d) is often represented by the largest of the three principal values, defining the ordering $S_{\parallel}^{(a)} \equiv \left\langle 3\left(\hat{e}_{\parallel}^{(a)}\cdot\hat{a}\right)^2 - 1\right\rangle / 2$ along the major principal axis, $\hat{e}_{\parallel}^{(a)}$, and the difference of the other two, in ascending order, defining the biaxiality $\Delta S_{\perp}^{(a)} \equiv (3/2)\left\langle \left(\hat{e}_{\perp'}^{(a)}\cdot\hat{a}\right)^2 - \left(\hat{e}_{\perp''}^{(a)}\cdot\hat{a}\right)^2\right\rangle$ with respect to the principal axes $\hat{e}_{\perp'}^{(a)}, \hat{e}_{\perp''}^{(a)}$. In this representation, the three principal axes $\hat{e}_{\parallel}^{(a)}$, $\hat{e}_{\perp'}^{(a)}, \hat{e}_{\perp''}^{(a)}$ simply constitute a relabelling of the axes $\hat{h}_p^{(a)}, \hat{l}_p^{(a)}, \hat{m}$ in ascending sequence of the respective principal values $S_{h_p h_p}^{(a)}$, $S_{l_p l_p}^{(a)}$ and $S_{mm}^{(a)}$.

***Vector-pseudovector order parameters.*** The local $C_2$ phase symmetry combined with the $C_{2V}$ molecular symmetry, give rise to nontrivial values for some components of the vector-pseudovector ordering tensors $S_{AB}^{*(a)}$. With the *y*-axis assigned as the $C_2$ molecular axis, these are defined as

$$S_{AB}^{*(a)} \equiv \frac{1}{2}\left\langle \left(\hat{a}\cdot\hat{A}\right)\left[\left(\hat{a}\times\hat{y}\right)\cdot\hat{B}\right] + \left(\hat{a}\cdot\hat{B}\right)\left[\left(\hat{a}\times\hat{y}\right)\cdot\hat{A}\right]\right\rangle, \quad (2)$$

Clearly $S_{AB}^{*(y)}$ is null and $S_{AB}^{*(z)} = -S_{AB}^{*(x)} \equiv S_{AB}^{*}$. For the local $C_2$ symmetry of the phase, the only non-vanishing components of the vector-pseudovector ordering tensor in the $\hat{h},\hat{l},\hat{m}$ frame are the diagonal elements $S_{hh}^{*}, S_{ll}^{*}, S_{mm}^{*}$ and the off-diagonal element $S_{hl}^{*}$. Due to the null sum of the diagonal elements, only two of them are independent. These are conveniently chosen to be $S_{hh}^{*}; S_{mm}^{*} - S_{ll}^{*}$. These describe correlations between the tilt of the molecular axes $\hat{x}$, $\hat{z}$ and the polar order of the molecular symmetry axis $\hat{y}$ in the local $\hat{h},\hat{l},\hat{m}$ phase axes. $S_{hl}^{*}$ includes, in addition, polarity-biaxiality correlations (See S2.3).

Determination of the $S_{AB}^{(a)}$, $\langle a_A\rangle$ and $S_{AB}^{*}$ components within a given slab at Z, provides the basic description of the local orientational ordering in a phase showing both polarity and chirality. The reorientation of the principal axes of these order parameters on moving from slab to slab constitutes the modulation. The ordering within a given slab is conveniently described with respect to the local, slab-fixed axes $\hat{h},\hat{l},\hat{m}$. The simulation data are analyzed in the context of the roto-translation symmetry predicted by the molecular theory of the $N_{PT}$ (*11*). This symmetry implies that on translating along Z, the directions $\hat{l},\hat{m}$ rotate about $\hat{h}$ by an angle $\varphi$ that is proportional to the translation, i.e. $\varphi = kZ$, with $k \equiv 2\pi/d$ the modulation's wavenumber, and $d$ its periodicity.

***Correlation functions.*** Under periodic boundary conditions, the absolute Z position of slabs in the simulation box is inherently arbitrary for uniform density, periodic phases. This interferes with the consistent determination of the Z-dependence of the orientations of the principal axes associated with local ordering. To overcome this, we calculate intermolecular orientational correlations, as these depend on relative coordinates and not on absolute slab positions. The calculation of the Z-dependence of these correlation function allows the determination of the local values of order parameters as well as the form of the modulation directly from the simulations. In connection with the three types of order parameters, $\langle a_A\rangle$, $S_{AB}^{(a)}$ and $S_{AB}^{*}$ defined above, we introduce the respective correlation functions, $g_1^{(yy)}(Z)$, $g_2^{(aa)}(Z)$ and $g^{*}(Z)$, for nematics with local $C_2$ symmetry showing roto-translational modulations in one dimension (chosen along the Z-axis) as follows.

The functions describing the correlation of polar ordering are defined by

$$g_1^{(ab)}(Z) = \left\langle \sum_{i,j}\left(\hat{a}(\mathbf{r}_i)\cdot\hat{b}(\mathbf{r}_j)\right)\delta(Z - \mathbf{r}_{ij}\cdot\hat{Z})\right\rangle \Big/ \left\langle \sum_{i,j}\delta(Z - \mathbf{r}_{ij}\cdot\hat{Z})\right\rangle \quad (3)$$

Here $\mathbf{r}_{ij} = \mathbf{r}_i - \mathbf{r}_j$ is the intermolecular vector between a pair of molecules $i, j$ positioned at $\mathbf{r}_i$ and $\mathbf{r}_j$, with their molecular axes in directions $\hat{a}$ and $\hat{b}$. $\delta$ denotes the usual Dirac delta-function.

As a result of the $C_{2V}$ symmetry of the curved-rod molecules in figure (2), only $g_1^{(yy)}(Z)$ is a non-vanishing function. Its evaluation from the simulation data is straightforward: For a given snapshot, two thin slabs perpendicular to $\hat{Z}$ and separated by a distance *Z* are chosen, the scalar products of the inter-slab pairs $\hat{y}_i\cdot\hat{y}_j$ are summed and divided by the total number of such pairs. The procedure is repeated for different slabs separated by the same *Z* and the averaging over all such pairs of slabs is obtained for the given snapshot. Further averaging is done for different snapshots for the same distance *Z* to obtain the calculated value of $g_1^{(yy)}$ at that *Z*.

The correlations of 2nd rank vector-vector ordering are conveyed by the functions:

$$g_2^{(ac;bd)}(Z) = \frac{\left\langle \sum_{i,j} \begin{pmatrix} \frac{3}{2}\left(\hat{a}(\mathbf{r}_i)\cdot\hat{b}(\mathbf{r}_j)\right)\left(\hat{c}(\mathbf{r}_i)\cdot\hat{d}(\mathbf{r}_j)\right) \\ -\frac{1}{2}\delta_{ac}\delta_{bd} \end{pmatrix} \delta(Z - \mathbf{r}_{ij}\cdot\hat{Z}) \right\rangle}{\left\langle \sum_{i,j} \delta(Z - \mathbf{r}_{ij}\cdot\hat{Z}) \right\rangle} \quad (4)$$

where $\hat{a}, \hat{b}, \hat{c}, \hat{d}$ denote molecular axis directions at the indicated positions. For molecules of $C_{2V}$ symmetry, the surviving functions are those with *a=c* and *b=d* of which only 3 are independent (See S3.1). These can be chosen to be $g_2^{(xx;xx)}(Z), g_2^{(yy;yy)}(Z), g_2^{(zz;zz)}(Z)$ and we denote them for brevity as $g_2^{(xx)}(Z), g_2^{(yy)}(Z), g_2^{(zz)}(Z)$ and can be obtained directly from the simulations in close analogy to $g_1^{(yy)}$.

Lastly, for the correlations of vector-pseudovector ordering, the appropriate function is (See S3.2):

$$g^*(Z) = \frac{\left\langle \sum_{i,j} \left( \left(\hat{z}(\mathbf{r}_i)\cdot\hat{z}(\mathbf{r}_j)\right) \begin{pmatrix} \left(\hat{z}(\mathbf{r}_i)\times\hat{y}(\mathbf{r}_i)\right)\cdot \\ \left(\hat{z}(\mathbf{r}_j)\times\hat{y}(\mathbf{r}_j)\right) \end{pmatrix} \right) \delta(Z - \mathbf{r}_{ij}\cdot\hat{Z}) \right\rangle}{\left\langle \sum_{i,j} \delta(Z - \mathbf{r}_{ij}\cdot\hat{Z}) \right\rangle}, \quad (5)$$

and can be directly obtained from the simulations as outlined for $g_1^{(yy)}$.

***Correlations according to the roto-translation modulation.*** Theoretical expressions relating all the above correlation functions to the local order paramaters can be readily obtained for the specific form of the modulation dictated by the rot-translation model. Thus we have (see S3.3) the following functional forms for their Z-dependence:

$$g_1^{(yy)}(Z) = p_\perp^2 \cos kZ , \quad (6)$$

relating the single polar correlation function to the magnitude of the polar order parameter $p_\perp$ and the wave number $k$ of rototranslation modulation .

Also,

$$g_2^{(aa)}(Z) = \left(S_{hh}^{(a)}\right)^2 + \frac{4}{3}\left(S_{lh}^{(a)}\right)^2 \cos kZ + \frac{1}{3}\left(S_{mm}^{(a)} - S_{ll}^{(a)}\right)^2 \cos 2kZ \quad , (7)$$

for $a = x, y, z$, according to which the vector-vector correlation functions are fully specified in terms of the respective order paramaters and the modulation wave number, and *vice versa*.

A similar expression is obtained for the "chiral" correlation function,

$$g^*(Z) = \left(S_{hh}^*\right)^2 + \left(\frac{4}{3}\left(S_{hl}^*\right)^2 + \frac{3}{4}\langle \hat{y}\cdot\hat{m}\rangle^2\right)\cos kZ + \\ +\frac{1}{3}\left(S_{mm}^* - S_{ll}^*\right)^2 \cos 2kZ \quad , \quad (8)$$

relating it to the magnitudes of the respective vector-pseudovector order parameters.

## 3. Simulation results and phase characterization.

### 3.1 Phase diagram.

Figure 3 shows distinct phases as a function of packing fraction *h*. Reduced pressure $P^*(\eta)$ equation of state (EoS) features imply that the system forms three phases: (i) isotropic, up to $P^* \approx 2.2$ ($\eta \approx 0.245$), which is the lower pressure limit of (ii), an orientationally ordered state that remains stable up to pressures of at least $P^* \approx 4.3$ ($\eta \approx 0.365$), the lower limit of stability of (iii) an orthogonal, biaxial smectic phase in expansion runs; in compression runs, above $P^* \approx 4.3$ the system shows substantial positional correlations.

### 3.2. Macroscopic and local ordering.

Following the standard principal axis analysis of the 2nd rank ordering tensors $S_{AB}^{(a)} = \langle 3a_A a_B - \delta_{A,B}\rangle / 2$, formed by ensemble-averaging the indicated combinations of the projections of the $a = x, y, z$ molecular axes on the $A,(B) = X, Y, Z$ axes of the simulation box, it is unambiguously found that for packing fractions in the range $0.244 < \eta < 0.37$, the system is <u>macroscopically</u> (here meaning over the entire extent of the simulation box) uniaxial with a unique axis, $\hat{h}$ (the box Z-axis).

In the same packing fraction range, on dividing the simulation box into a stack of consecutive slabs of thickness $D$ (rod diameter) perpendicular to the macroscopic symmetry axis, $\hat{h} \parallel \hat{Z}$, it becomes clear that:

1) the average number of molecules per slab $N(Z)$ is independent of $Z$, indicating positional uniformity.

2) the polar order parameter of the ordering of the *y*-molecular axis, $\mathbf{p}(Z)$, is non-zero and is directed perpendicular to $\hat{h}$. Its direction varies with Z while its (pressure-dependent) magnitude, $p_\perp$, is independent of Z. With $\hat{m}(Z)$ a unit vector perpendicular to $\hat{h}$ and coincident with the direction of the polar ordering within a slab, one may set $\mathbf{p}(Z) = p_\perp \hat{m}(Z)$; i.e., $\hat{m}(Z)$ is the direction of maximum alignment of the *y*-molecular axis. To complete a local frame, the unit vector $\hat{l}(Z)$ is defined perpendicular to both $\hat{h}$ and $\hat{m}(Z)$.

3) The eigenvectors of the ordering tensors $S_{AB}^{(a)}(Z), S_{hh}^*(Z)$ in each slab, which define the directions of the tensor principal axes, exhibit a clear, periodic *Z*-dependence; the eigenvalues are independent of $Z$. The directions of maximum alignment of the molecular *z*- and *x*-axes are neither parallel nor perpendicular to $\hat{h}$, remaining on a plane containing the latter and perpendicular to $\hat{m}(Z)$ (see inset fig 2d); that plane rotates about $\hat{h}$ on translating along the Z direction.

4) The per-slab analysis shows that among the several principal axes associated with local ordering, the polar direction $\hat{m}(Z)$ is the single common principal axis to all ordering tensors defined in a given slab; this identifies $\hat{m}(Z)$ as the unique local symmetry axis of the phase.

Extracting ensemble averages from the entire simulation box, while not providing detailed information on the local molecular ordering, is useful for identifying the presence and general features of global ordering, singling out the directions associated with that ordering, and for monitoring possible phase transitions. To illustrate this point we show, on the same diagrams of Fig. 4 the evolution of the Z-averaged components of the pertinent ordering tensors $S^{(a)}_{AB}(Z), S^{*}_{hh}(Z)$ as the system moves through transformations between the isotropic, the modulated nematic and the layered phase, both in expansion and compression sequences.

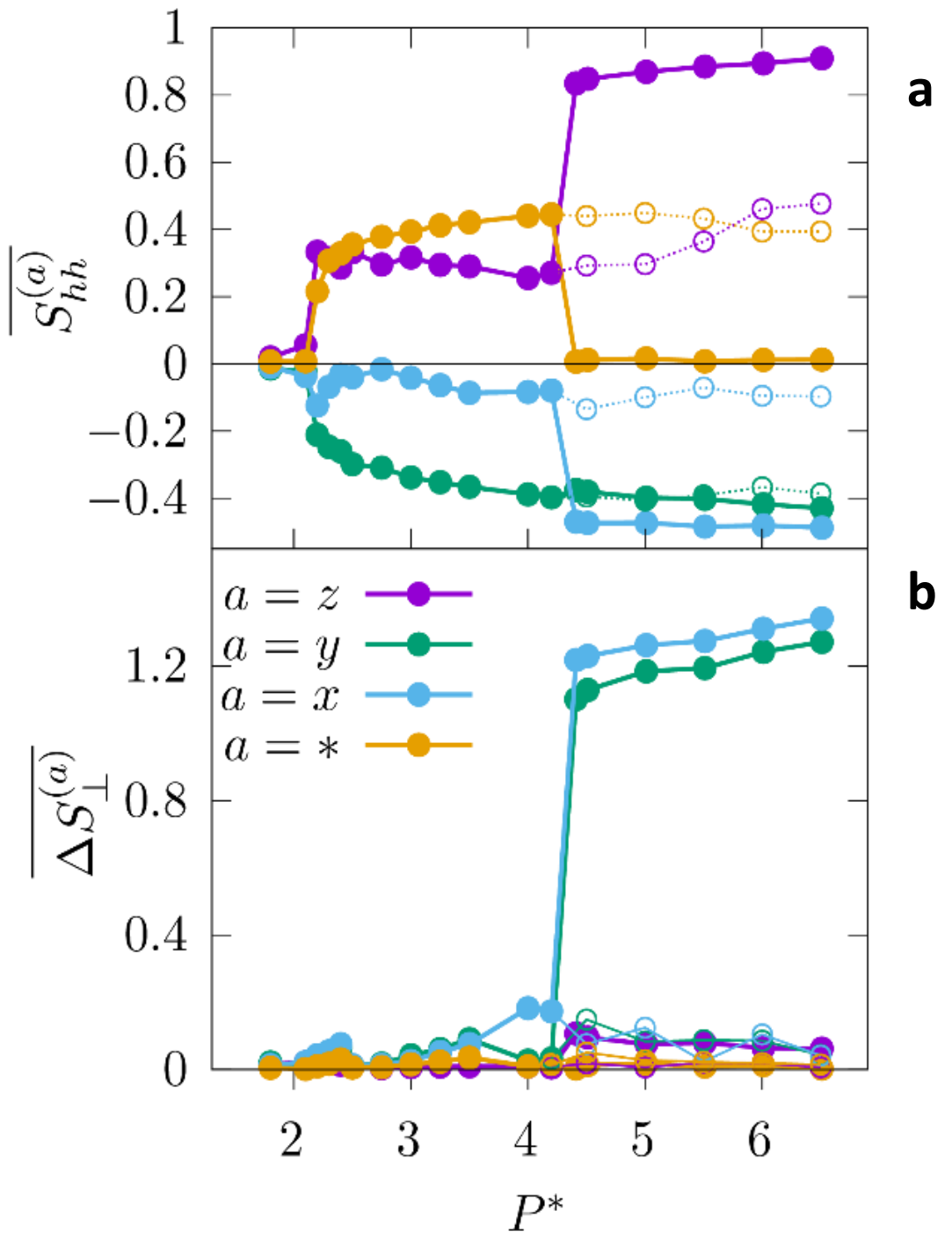

**Figure 4 Pressure dependence of the macroscopic (spatially averaged) order parameters** (a) and their biaxialities (b), associated with various averages of the indicated ordering tensor components over the simulation box. The filled symbols correspond to expansion runs from the high-pressure smectic phase and the open symbols are for compression runs from the $N_{PT}$ phase.

The density dependence of the nematic order parameter associated with ordering of the *z*-molecular axes with respect to $\hat{h}$, $\overline{S^{(z)}_{hh}}$, (Fig. 3 and Fig. 4a), shows unconventionally low values which decrease further with increasing density. The system is also macroscopically nonpolar, i.e., null values of the first rank ordering tensors $\langle a_A \rangle$ on averaging over the entire simulation box. Notably, the system shows macroscopic chirality—a nonvanishing value of $\overline{S^{*}_{hh}}$. Within the range $0.244 < \eta < 0.37$, all ordered systems studied are nematic since they only present short-range mass-density correlations.

In the modulated nematic phase, the modulation axis, $\hat{Z} \parallel \hat{h}$, is determined by the eigenvector associated with the maximal eigenvalue of $\overline{S^{(z)}_{AB}}$. It then follows that $\overline{S^{(y)}_{hh}}$ has negative values, see fig. 4a, implying that the polar *y*-axis of the molecules is preferentially ordered perpendicular to the modulation axis. It also follows that $\overline{S^{(x)}_{hh}}$ is small in magnitude but persistently negative and, for all three molecular axes, the structure of the calculated $\overline{S^{(a)}_{AB}}$ is practically uniaxial, in accord with the theoretical result in eq(S19). Shown in figure 4a are also the values of $\overline{S^{*}_{hh}}$, extracted directly from the simulation data with the same procedure of averaging over the entire simulation box and showing the same uniaxial structure in accord with the theoretical result. The non-vanishing value of $\overline{S^{*}_{hh}}$ reflects directly the presence of macroscopic structural chirality. As expected, averaging over the entire box yields for the polar order parameters $\overline{p_X} = \overline{p_Y} = \overline{p_Z} = 0$ .

Aside from the usual abrupt changes across the isotropic to nematic transitions, the following features are notable on the high-pressure side of the diagrams in Fig. 4:

(i) The layered phase is clearly biaxial on the global scale, as shown by the non-vanishing values of the difference $\overline{\Delta S^{(a)}_{\perp}} = \overline{S^{(a)}_{XX}} - \overline{S^{(a)}_{YY}}$ (and similarly $\overline{\Delta S^{*}_{\perp}} = \overline{S^{*}_{XX}} - \overline{S^{*}_{YY}}$ ) between the *Z*-averaged ordering along the two eigen-vectors (defining the macroscopic axes $\hat{X}, \hat{Y}$ for each *(a)* or ***) perpendicular to the primary eigenvector $\hat{Z} \parallel \hat{h}$ defining the macroscopic principal axis of the layered medium.

(ii) The negligible biaxiality of $\overline{\Delta S^{(y)}_{\perp}}$ implies that, although the layered phase is biaxial and the polar molecular *y*-axis has a relatively high quadrupolar order along its primary axis of ordering, its orientational fluctuations in the transverse directions show no strong preference for one of the axes over the other.

(iii) The vanishing values of $\overline{S^{*}_{hh}}$ in the layered phase indicate the absence of any persistent structural chirality, at least over the length scale of the simulated layered systems.

In summary, the results of extracting the spatially averaged components of the ordering tensors directly from the simulation show that the modulated nematic phase is macroscopically apolar, uniaxially ordered and structurally chiral. It is emphasised that a macroscopic sample is understood here to extend over the volume of the simulation box, wherein we get a single chiral domain of definite handedness.

### 3.3 Direct calculations of local ordering from simulation data.

The Z-dependence of $g^{(yy)}_{1}(Z)$ is shown in Fig. 5a together with the form implied by the roto-translation formulation of the modulation according to eq(6). It is apparent that this prediction agrees perfectly with the Z-dependence obtained from simulations. This provides the respective values, shown in figure 6, for the polar order parameter $p_{\perp}$ and the modulation wave vector $k$ at each pressure. Similarly, the three independent components $g^{(xx)}_{2}(Z)$, $g^{(yy)}_{2}(Z)$, $g^{(zz)}_{2}(Z)$ and the “chiral” correlation function $g^{*}(Z)$, are shown in fig. 5. The functional forms include a constant term and two harmonics, $\sim \cos kZ$ and $\sim \cos 2kZ$ . These are perfectly reproduced by the theoretical predictions of the roto-translation model of the modulation (*11*), allowing, according to eqs 7 and 8, the evaluation of the optimal values of the full set of 2nd rank order parameters plotted in fig. 6b-d and a value for the modulation wave vector *k* that coincides with the values obtained from the respective fittings with the expressions for the polar correlation function $g^{(yy)}_{1}(Z)$ .

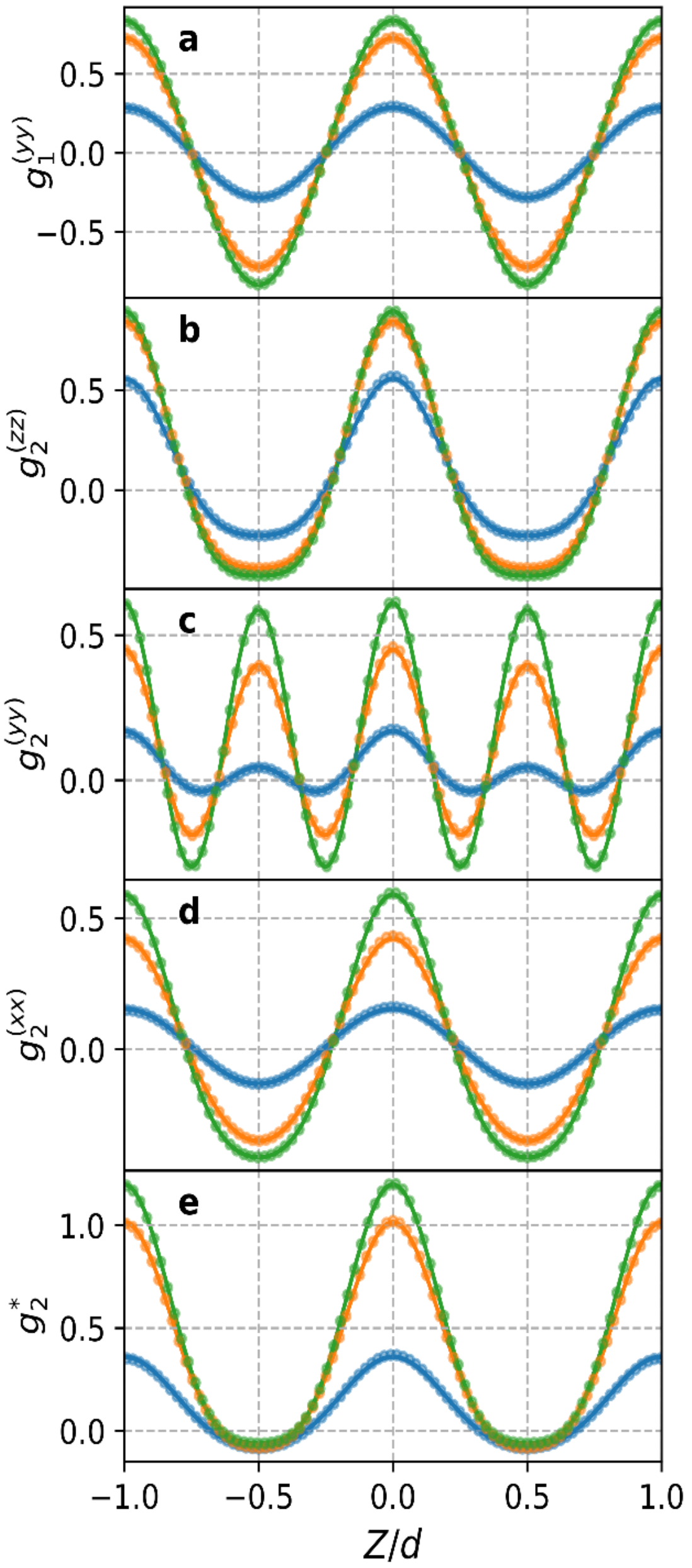

**Figure 5 First and second rank orientational pair correlations**. Computed $g_1^{(aa)}(Z), g_2^{(aa)}(Z)$ and $g^*(Z)$, at three different reduced pressures in the modulated nematic phase covering the full range of its stability: at $P_1^* = 2.2$ (blue) which is close to N-I phase transition pressure, at $P_2^* = 3.0$ (orange) deep in the modulated nematic phase and at $P_3^* = 4.0$ (green) which is slightly below the pressure at which the smectic phase becomes unstable upon expansion of the system. These pressures correspond to packing fractions $\eta_1 = 0.244$, $\eta_2 = 0.303$ and $\eta_3 = 0.348$, respectively. The symbols are results obtained directly from the simulations and the solid lines are fits based on the theoretical roto-translation model.

The ordering of the three molecular axes (fig. 2(d)) described by the order parameters $S_{AB}^{(a)}$, is conveniently analyzed by specifying:

- the direction (major principal axis) of maximum alignment $\hat{e}_{\parallel}^{(a)}$;
- the value of the order parameter along that axis $S_{\parallel}^{(a)} \equiv \left\langle 3\left(\hat{e}_{\parallel}^{(a)} \cdot \hat{a}\right)^2 - 1\right\rangle / 2$;
- the directions of the two minor principal axes, $\hat{e}_{\perp'}^{(a)}, \hat{e}_{\perp''}^{(a)}$, perpendicular to $\hat{e}_{\parallel}^{(a)}$;
- and the value of the biaxiality, $\Delta S_{\perp}^{(a)} \equiv (3/2)\left\langle \left(\hat{e}_{\perp'}^{(a)} \cdot \hat{a}\right)^2 - \left(\hat{e}_{\perp''}^{(a)} \cdot \hat{a}\right)^2 \right\rangle$.

These are readily obtained from the order parameters in fig. 6b-d. The results are shown in fig. 6e-g. The polar molecular y-axis orders, by definition, along the symmetry axis $\hat{m}$. Fig.6e shows that the principal value of the respective ordering tensor is relatively high, reaching $S_{\parallel}^{(y)} \approx 0.8$ at high pressures. The other two principal axes are obtained on rotating by $\theta^{(y)}$; this ranges from about 40º to 46º, (see fig. 6g) increasing with pressure and yielding a decreasing biaxiality $\Delta S_{\perp}^{(y)}$ from about 0.4 to 0.2 (fig. 5f). Analogous trends and numerical values are obtained for the x-molecular axis, except that the maximum ordering is along the rotated $\mathbf{e}_{\parallel}^{(x)}$ axis, obtained on rotating about $\hat{m}$ by $\theta^{(x)}$; the latter persistently exceeds $\theta^{(y)}$ by 1-2º. The maximum alignment of the z-molecular axis is along the $\hat{e}_{\parallel}^{(z)}$ axis, obtained on rotating by $\theta^{(z)}$ (often called the "tilt" or "cone" (*36*) angle of the "director" in the erroneous $N_{TB}$ terminology), which is persistently smaller than $\theta^{(y)}$ by 2-3º. The degree of alignment $S_{\parallel}^{(z)}$ is considerably higher than $S_{\parallel}^{(y)}$ and $S_{\parallel}^{(x)}$. Notably, the biaxiality $\Delta S_{\perp}^{(z)}$ in the principal axis frame is nearly vanishing over the entire pressure range of the modulated nematic phase (fig. 6f).

The $S_{hl}^*$ order parameter, a measure of the polarity-biaxiality correlations, (see S2.3), remains slightly below zero in the modulated nematic phase, see fig. 6(d). The calculated values of the rotation angle $\theta^*$ that diagonalises the $S_{AB}^*$ ordering tensor (Fig. 7) are relatively small right below the I-N transition and further decrease to a marginal magnitude with increasing pressure, indicateing the diminishing of polarity-biaxiality correlations deep in the nematic phase.

## 4. Discussion

Distinguishing macroscale form chirality in the locally uniaxial ferroelectric nematic ($N_F$) and twist-bend nematic ($N_{TB}$) from nanoscale modulations in the polar-twisted nematic ($N_{PT}$) is possible by measuring orientational pair correlations. Our simulations reveal non-vanishing values of the "polar" order parameter $p_{\perp}$ and the "chiral" order parameters, $S_{hh}^*, S_{mm}^* - S_{ll}^*, S_{hl}^*$: these would strictly vanish for local uniaxial and apolar ordering, i.e., in $N_U$ and its elastically modulated forms, for example, the $N_{TB}$. Moreover, the structure of $S_{AB}^{(a)}$ is precisely that of a medium with local $C_2$ ordering featuring a polar director $\hat{m}$. This is in sharp contrast with the uniaxial apolar ordering about the nematic director $\hat{n}$ inherent to the $N_U$ and the $N_{TB}$. In addition, the $N_{PT}$ modulation wavenumber k is two orders of magnitude larger than any wavenumber that could accommodate elastic twist or bend deformations of $\hat{n}$. The marginal local biaxiality of $S_{AB}^{(z)}$ should not lead to mistaking the principal axis $\hat{e}_{\parallel}^{(z)}$ for a symmetry axis nor for a nematic director $\hat{n}$, as there is very strong polar ordering along $\hat{m}$ (i.e.

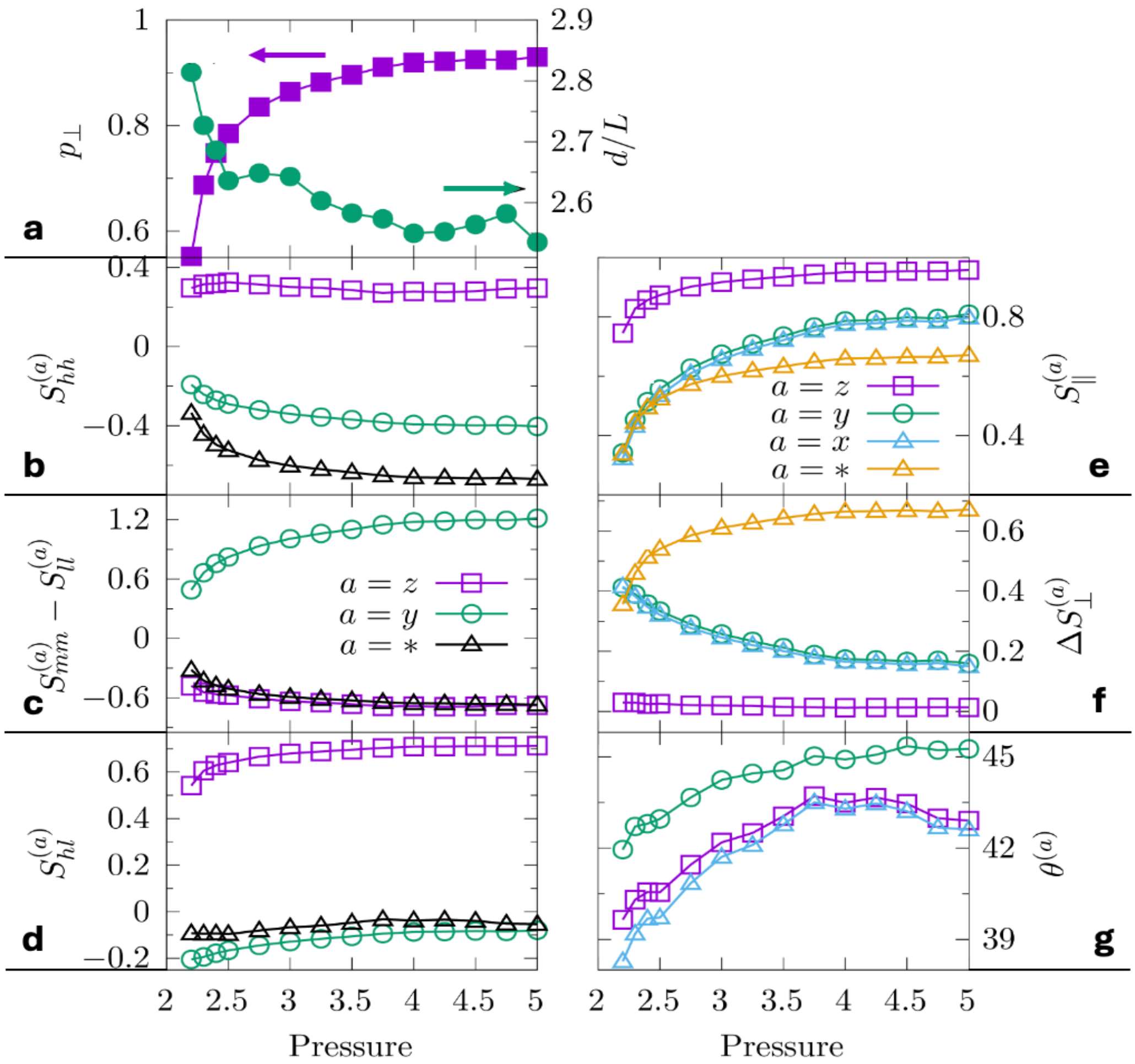

**Figure 6 Pressure dependence of order parameters.** Optimal values of: **a** polar order parameter (squares) and periodicity length (circles) in units of the molecular length, **b-d** second rank order parameters, evaluated by fitting the correlation functions in fig. (4); **e-f** principal values of the ordering tensors and **g** of the corresponding rotation angles (their sign inverts with the handedness of the helical modulation), for the three molecular axes *x,y,z* determined from the order parameter values shown in **b-d**.

transverse to $\hat{e}_{\parallel}^{(z)}$ ). Furthermore, the maximum alignments of the $\hat{x}$ and $\hat{y}$ molecular axes are along distinctly different principal axes, and the respective ordering tensors are highly biaxial. We stress this because in previous simulations (*36*, *37*) the $S_{AB}^{(z)}$ tensor was used as the sole measure of the ordering in modulated nematic phases, and its marginal biaxiality could incorrectly be taken to suggest that what twists along the modulation axis is a tilted uniaxial nematic director $\hat{n}$ , as would indeed be the case in a true $N_{TB}$.

Clearly, the very presence of $\hat{m}$ as a symmetry axis excludes the possibility of local $D_{\infty h}$ symmetry. Furthermore, the inequivalence of $\hat{m}$ and $-\hat{m}$ (i.e., locally polar ordering) implies that the only symmetries compatible with the 2nd rank tensorial properties correspond to the $C_2$ or the $C_{2V}$ point groups with the $C_2$ axis along $\hat{m}$ . However, the identification of different angles, $\theta^{(x)}, \theta^{(y)}, \theta^{(z)}$ implies that no plane containing the $\hat{m}$ -axis can be a symmetry plane for the respective 2nd rank $S_{AB}^{(a)}$ and, therefore, $C_{2V}$ symmetry is ruled out. Thus, the only (local) symmetry element, common to all ordering tensors, is the polar $\hat{m}$ -axis, implying a locally $C_2$ polar nematic. This structural picture has already been corroborated by a simple molecular model (*11*), and subsequently confirmed by molecular dynamics simulations(*12*).

Our simulations using curved rods show that the nanoscale twist-bend nematic proposal based on spontaneous elastic twist and bend deformations is incorrect, and moreover, that the smectic-like phases reported in ref. 32 are artifacts stemming from improper simulation box dimensions. Our results are in agreement with that earlier work for the packing fraction at the N-I transition. However, there is marked disagreement regarding (i) the interpretation of the structure and (ii) the phase sequence at higher packing fractions. In particular: (i) The roto-translational modulation of curved rod simulations has nothing to do with the elastic twist or bend deformations found in common $N_U$ LCs. Detailed symmetry and order of magnitude arguments have been presented (*33*, *34*, *45*, *46*), and confirmed herein; the modulation length scale, strong polar ordering, and symmetries of the ordering tensors place the observed modulation well outside the validity of nemato-elasticity. (ii) The simulations in Chiappini & Dijkstra (*36*) report the sequence I-$N_{TB}$-$Sm_{SB}$-$Sm_{TSB}$ (smectic phases with splay-bend and twist-splay-bend deformations, respectively). The authors also clearly state that the different phases ($Sm_{SB}$, $Sm_{TSB}$ and $N_{TB}$,) for the $L/D = 19$ , $\Psi = 1.31$ system have been obtained in simulations with one box side fixed $L_x = 30.3D$ ($Sm_{SB}$), 33.4D ($Sm_{TSB}$), and 36.3D ($N_{TB}$). These lengths, however, are well below the $37.4D$ molecular length threshold required by this particular system (see section S.1 in SIF) for the exclusion of ghost interactions. This is a significant flaw that renders their results, especially those referring to phase transitions, unreliable even at relatively low packing fractions. In fact, in very long simulations at the packing fractions corresponding to the presumed $Sm_{SB}$ and $Sm_{TSB}$ in (*36*), we find that the stable phase is $SmA_BP_A$, an orthogonal biaxial smectic phase consisting of polar layers with an antiferroelectric arrangement (Fig. 3); on lowering the packing fraction η, this stratified structure transforms into the helicoidal polar-twisted nematic. The absence of the so-called $Sm_{TSB}$ is also confirmed by other simulations of similar rigid-core models (*43*).

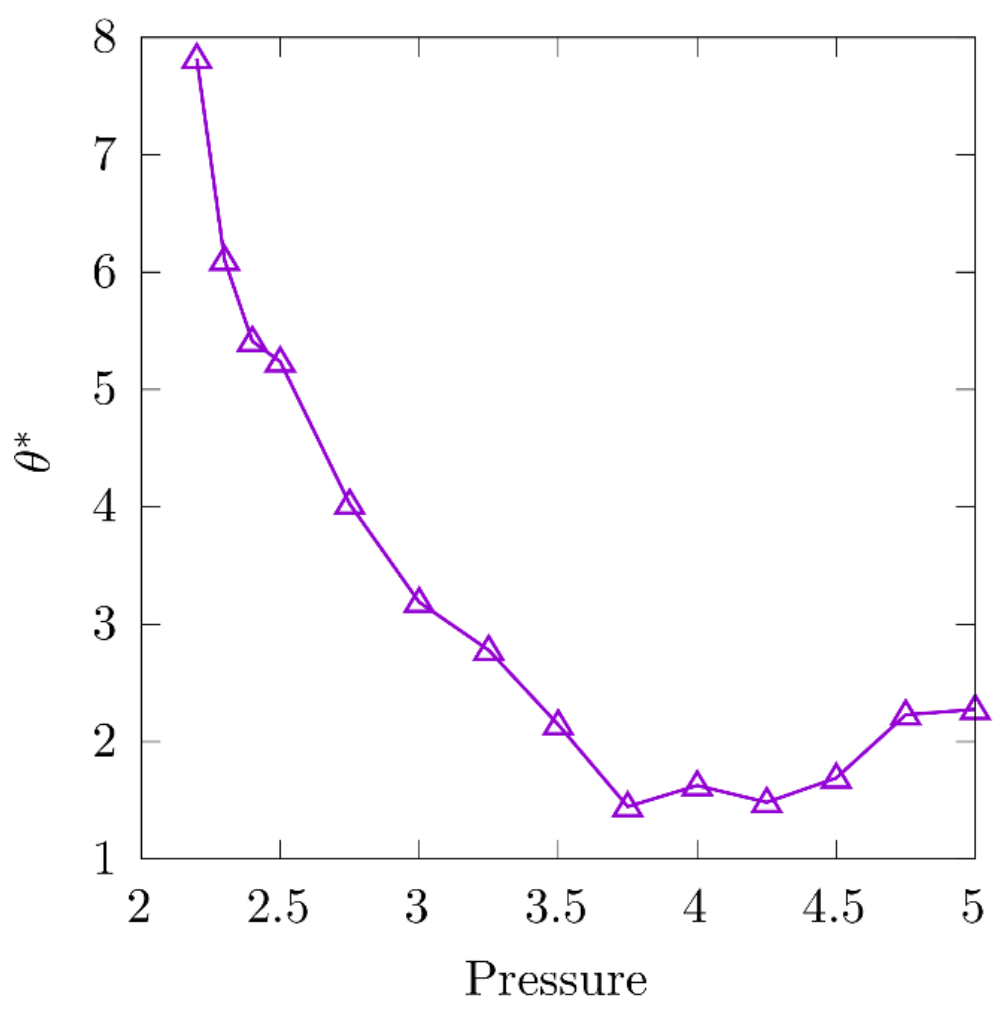

**Figure 7 Calculated pressure dependence of the rotation angle** $\theta^*$ that diagonalises the $S^*_{AB}$ ordering tensor in the modulated nematic phase range.

We note that starting from strictly linear rods (Ψ=0) the well-known Onsager I- $N_U$ transition occurs at a critical rod density. On increasing rod curvature up to Ψ~0.5 we still get direct I - $N_U$ transitions without a $N_{PT}$ phase. Above Ψ~0.5 the $N_{PT}$ appears and we get the phase sequence I - $N_U$ - $N_{PT}$ with increasing density. The rod density range of the $N_U$, however, decreases with increasing Ψ. At about Ψ=1.2 the $N_U$ disappears, and we find direct I - $N_{PT}$ transitions for Ψ>1.2. This was meticulously demonstrated by Chiappini & Dijkstra (*36*) and agrees with our replication of their phase transitions, but with a totally different interpretation of the origins of the nano-modulation, and with an artifact-free smectic phase that transforms into the nano-modulated nematic.

The misidentification of the "$N_X$ phase" of CB7CB dimers as a twist-bend (TB) nematic continues to propagate in the literature. Its popularity is so widespread that one group of investigators of the newly-discovered spontaneous mirror symmetry breaking in the $N_F$ phase advocates analogies with the 8-nm modulated phase of CB7CB even though the $N_F$ has a helical structure with a pitch in the range of visible light wavelengths; Karcz et al. inflate the "twist-bend family" with another classification, $N_{TBF}$ (*9*). The wide-ranging TB designation is further extended in *ad hoc* modelling of the $N_X$ phase as a 3D fluid displaying a nanoscale helical pitch and consisting of a self-assembled hierarchy of duplex helices comprised of entertwinded, concatenated CB7CB dimers (*47*). We find no evidence of such helical pairing in our simulations.

Meyer's elegant continuum description of the $N_{TB}$ and his transparent nomenclature is unambiguous. He asks, "Is there a space filling structure containing uniform bending of the director field?" and answers, "There is; if one starts with a cholesteric structure, and adds a constant component of the [uniaxial nematic] director parallel to the helix axis, the result is a state of uniform torsion and bending, the relative amounts of bend and twist being determined by the tilt angle of the director relative to the helix axis."(*48*)

A physically sound understanding of the origins of form chirality appears to be a prerequisite for understanding spontaneous breaking of mirror symmetry, i.e., forming chiral structures with achiral mesogens. In both the $N_{TB}$ and $N_F$ phases, equilibrium *macroscale* deformations of $\hat{n}$ result from a delicate energetic balance: chiral trajectories of $\hat{n}$ decrease electrostatic energy—flexoelectric and dipolar, respectively—which in turn, is offset by an increase in inelastic energy. By contrast the *nanoscale* modulations characterizing the $N_{PT}$ phase have local $C_2$ symmetry and, as a result of this symmetry, reveal a unique director for the phase that lacks the typical symmetries of the nematic director $\hat{n}$ . Hence, that modulation is not amenable to the Frank-Oseen continuum-elasticity description used by Meyer to conjecture splay- and twist-bend nematic structural polymorphs. Inappropriate application and casual dissemination of Meyer's cogent concepts into the vernacular obscure his concise descriptions of space-filling nematic director trajectories hampering the identification of new nematic polymorphs.

The results reported here reinforce the validity of the proposed polar-twisted nematic phase (*11*) confirming that the $N_{PT}$ is not a compressed version of Meyer's twist-bend nematic. The nanoscale form chirality in the $N_{PT}$ derives from achiral, polar-shaped dimer mesogens adopting a tight, chiral, packing motif—the roto-translation of the polar director $\hat{m}$ —and it is stabilized by the avoidance of a low-entropy, uniformly-polar (ferroelectric), fluid nematic phase. This, however, does not preclude the possibility of elastic deformations, on a much larger length scale, of the modulation axis $\hat{h}$ generating the analogues of twist, bend and splay distortinos of that axis, see ref. (*38*).

## 5. Conclusions

Resolving the structure and origins of the lower temperature polymorph in the odd-n linked CBnCB and related dimers has been a three-decades-long odyssey: First observed in 1991 by Toriumi et al(*30*), it was said to be a smectic phase in 1993(*49*). Then in 2011(*50*), after its definitive identification as a nematic, it was classified as Meyer's long-sought-after twist-bend nematic (*28*), despite its incommensurate elastic modulation scale. In 2013 the nanomodulated helical structure was confirmed experimentally(*31*, *32*). Finally in 2016 a molecular theory for the nematic-nematic transition was presented(*11*) arguing that the lower temperature nematic is a new nematic polymorph, the polar-twisted nematic, $N_{PT}$. However, since the 2011 assignment most researchers have continued to describe the phase as $N_{TB}$.

The nanoscale modulated structure is reproduced and interpreted consistently in our simulations of hard curved-shaped particles using the exact same parameterization used by others (*36*). It is a straightforward extension of Onsager's seminal description of entropy-stabilized nematic phase. When his linear rod-like particles are generalized to include rod curvature, additional constraints and symmetry considerations come into play: excluded volume packing dictates the formation of polar arrangements. But this additional polar order is antithetical to an entropy-stabilized structure. Consequently, the polar director $\hat{m}$ spirals into tightly wound helicoidal domains of left- and right-handed senses. On a macroscopic scale the blend of chiral domains appears to be a uniaxial and apolar nematic. Insofar as excluded volume interactions dominate electrostatic ones in liquid crystal structure formation, and insofar as our coarse grain model of dimers (Fig. 2c) captures the nonlinear mesogen architecture of odd-n CBnCB dimers, the simulation implications are incontrovertible: the polar mesogen shape drives the form chirality observed in the initially-designated $N_X$ phase of bent-core CBnCB dimers. While there is also an electrostatic contribution from the locally aligned dimers, i.e., the weak polarity associated with projecting the configurationally averaged, cyanobiphenyl dipoles on $\hat{m}$, that electrostatic component merely passively follows the excluded volume interactions that cause the nanoscale modulated $N_{PT}$ structure. In summary, the curved-rod simulations fully support that the second, lower temperature nematic phase in odd-n linked CBnCB dimers is a new nematic polymorph described by the polar-twisted nematic model.

## Conflicts of interest

There are no conflicts to declare.

## Acknowledgements

We thank Andrey Dobrynin and Daphne Klotsa for helpful discussions, and the initial referees who prompted a more emphatic explication.

## Notes and references

# Entropy stabilized form chirality in curved rod nematics: structure and symmetries.

*Alexandros G. Vanakaras[1], Edward T. Samulski*[2], and Demetri J. Photinos[1]*

[1] Department of Materials Science, University of Patras, Patras, 26504, Greece

[2] Department of Chemistry and Applied Physical Sciences, University of North Carolina, Chapel Hill, NC 27599-3201 USA

# Supplementary Information

### S.1 Simulation Details.

We have applied constant pressure Metropolis Monte Carlo simulations of systems composed of $N = 4096$ particles. Systems of half and/or double size were also simulated at certain pressures to study system size effects. On average, during a MC cycle, a random rotation/translation is attempted for each of the particles, chosen in a random fashion, as well as a random change of the system's volume is attempted. Random particle moves are accepted if they do not result to overlaps between the particles. Similarly, free of overlaps volume changes, $\Delta V = V_n - V_o$, are accepted with probability $\min\left(1, \exp[-P^* \Delta V + N \ln(V_n / V_o)]\right)$.The volume changes are performed allowing the lengths of the simulation box sides to fluctuate independently.

The size of the periodically modulated systems in these simulations should be large enough to ensure that at least one side of the simulation box is an integer multiple of the inherent periodicity length of system. Furthermore, none of the box lengths should be less than twice the cut-off distance of the intermolecular potential. For hard bodies this is given by the distance between the two most distant points on the surface of the particle. Accordingly, the cutoff distance $d_c$ for a pair of curved hard rods used in the simulations is $d_c / D = 1 + 2\frac{L}{D}\frac{\sin(\Psi/2)}{\Psi}$. For $L/D = 19$ and $\Psi = 1.31$ one obtains $d_c = 18.67D$. Therefore, none of the simulation box sides can be less than $2d_c \approx 37.4D$, otherwise ghost interactions may creep in as a given particle will be allowed to interact simultaneously with a second particle and with the periodic image of that same particle. This in turn introduces fictitious (self-ordering) fields that could bias severely or inhibit the accessible pathways connecting phases with substantially different types of ordering, as is the case in the present systems.

Denoting by $\sigma(\hat{r}_{12}, \omega_{12})$ the contact distance (i.e. the smallest distance before they start to overlap) between two hard particles having relative orientation, $\omega_{12}$ and direction of their relative positions along $\hat{r}_{12}$, $d_c$ is simply the maximum value of $\sigma(\hat{r}_{12}, \omega_{12})$. Under thermodynamic conditions of high orientational ordering of a hard body system, where the orientations of the individual particles exhibit a narrow distribution about a certain direction, $d_c$ may be obtained for interparticle position vectors also narrowly distributed along a given direction. In these cases, a carefully chosen simulation box-size, with only one box-side larger than $d_c$, could correctly simulate the system under the specific thermodynamic conditions. However, any phase transition resulting in a different distribution of the molecular orientations should be treated very cautiously. The simulations presented here are free of box-size conflicts with $d_c$ and the accompanying fictitious fields.

Several structures of highly packed crystalline states were used as the initial, high-pressure state. Expansion runs from the highly ordered close packed states were performed initially up to the melting point at which the positionally ordered phase melts into a positionally disordered state. This new state is then used as the initial state for further expansion runs to locate other possible phase transitions. The obtained pressure vs packing fraction equation of state, presented in the Figure 3 of the main text, indicates that in the reduced pressure range $P^* = 1-7$ the system exhibits three distinct phases, with isotropic, nematic and smectic-like structures.

Typical snapshots of the three phases are given in Fig. SI.1. We note here that compression runs from the isotropic phase suggest that an ordering transition is taking place at pressures slightly higher than the pressure $P^*_{NI}$ found in the expansion runs. The structure of the first ordered phase during compression runs is highly defective and depends strongly on the size and relative anisotropy of the sides of the simulation box. Similarly, compression runs starting from the modulated nematic phase do not show a sharp phase transition to a smectic phase. Instead, positional correlations develop gradually, and the system organizes into a fragmented phase with a high degree of local positional and orientational order (see Fig SI.1(d)). As this work is focussed on the structure and symmetries of the nematic phase, extensive studies on the structure of the highly ordered states are not presented here.

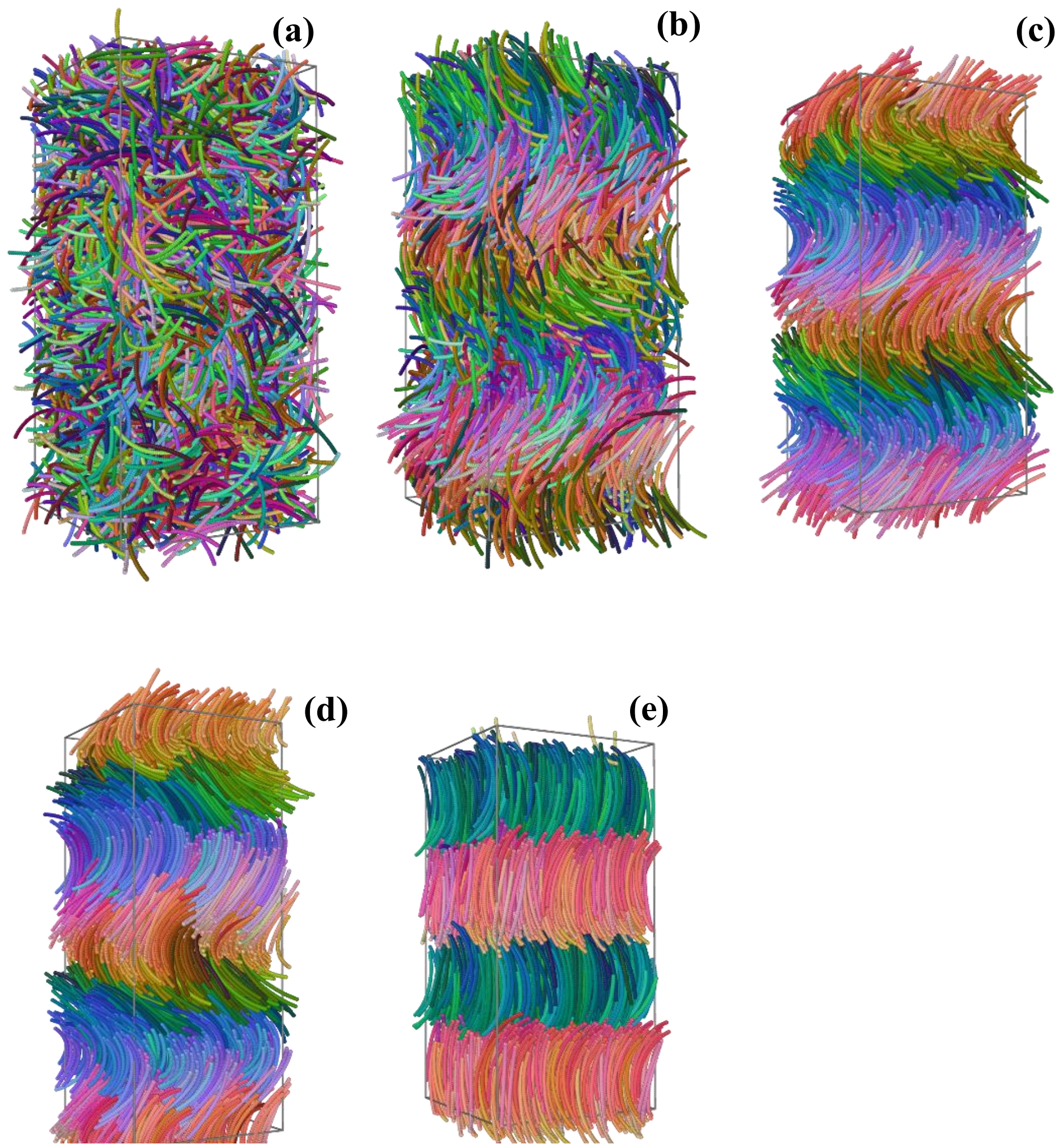

**Figure SI.1** Simulation snapshots: (a) Isotropic state at $P^* = 2.15$, (b) Modulated nematic at $P^* = 2.25$ and at $P^* = 4.00$ (c); (d) super compressed fragmented state at $P^* = 5.00$ obtained by continuous compression of a modulated nematic state; (e) Smectic state at $P^* = 5.00$ from expansion runs. The colouring of the particles is associated with different orientations of the molecular y-axis. The RGB colouring scheme has been chosen as $[R, G, B] = 0.5 + [y_X, y_Y, y_Z]/2$, with, $y_A$ the projection of the polar molecular axis along the sides of the simulation box, $0 < R, G, B < 1$.The length, in units of *D*, of the sides of the simulation box fluctuate slighgthly about the $(L_x, L_y, L_z)$ values: (a) (50.5, 52.8, 110.2); (b) (49.6, 49.3, 106.5); (c) (42.9, 43.6, 97.2); (d) (41.2, 41.3, 96.3); (e) (42.3, 42.8, 88.5)

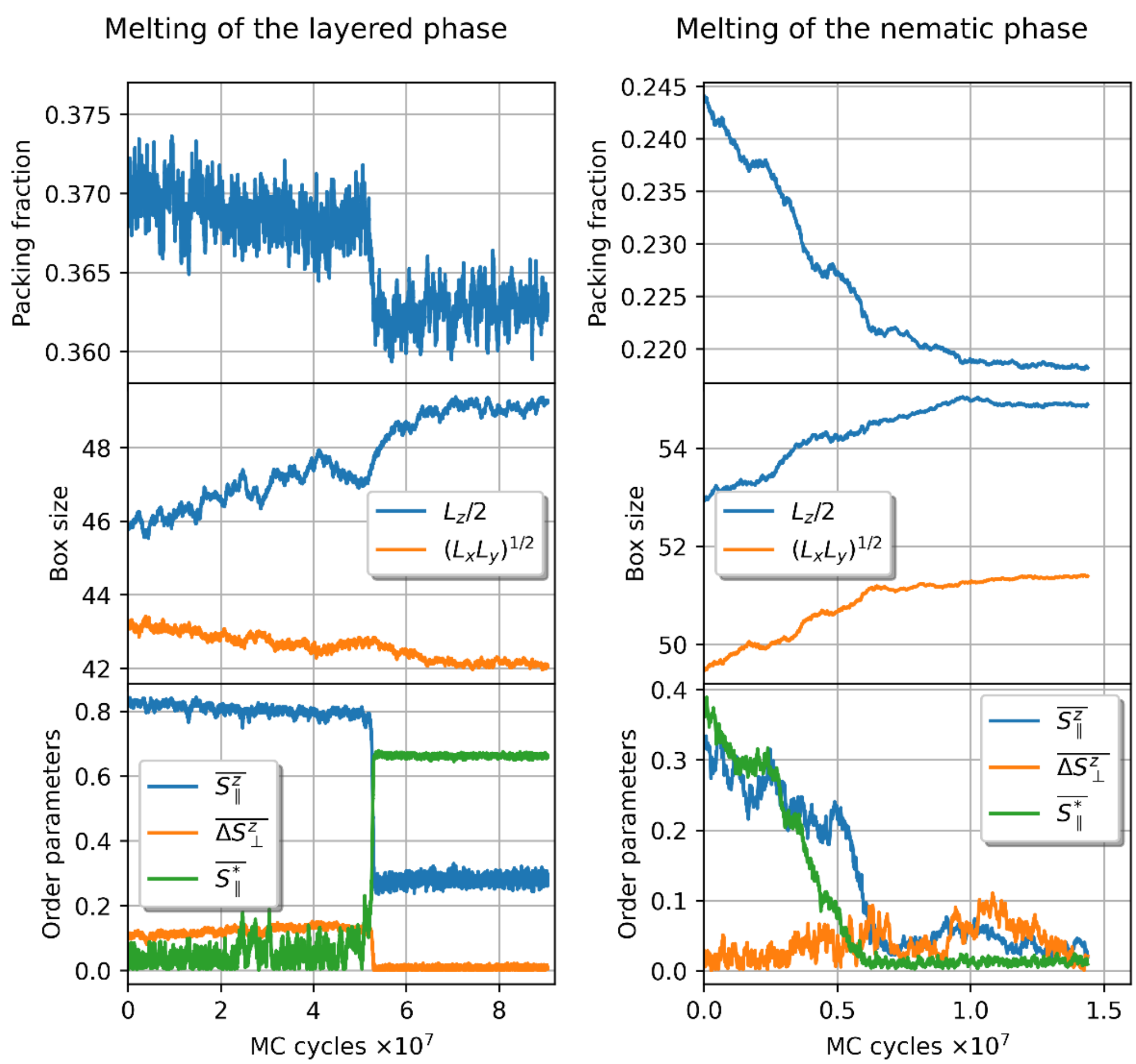

**Figure SI.2.** Evolution of the Smectic to Nematic and the Nematic to Isotropic phase change as function of the MC-cycles. The initial states are equilibrated at pressures for which the smectic (left) and the nematic state (right) are stable. These pressures are then reduced by $\Delta P^* = 0.05$ and the systems are left to evolve to their equilibrium.

In Fig. SI.2 we present the evolution of the packing fraction, the size of the simulation box and the main orientational order parameters along the smectic to nematic (left panel) and across the nematic to isotropic (right panel) phase transitions. $P^* = 4.40$ is the lowest simulated pressure for which a mechanically stable smectic phase appears, even after very long runs ($\sim 10^8$ MC-cycles), having an average packing fraction $\langle \eta \rangle \simeq 0.371$. On reducing the pressure to $P^* = 4.35$ the system reduces gradually its packing fraction down to $\langle \eta \rangle \simeq 0.368$ while preserving its smectic structure. Below this packing fraction, estimated as the lowest density for which the smectic phase is mechanically stable, the system melts into a modulated nematic phase showing an average packing fraction $\langle \eta \rangle \simeq 0.363$.

In the smectic phase the box accommodates four smectic layers with layer spacing $\sim 1.22L$ ; in the nematic phase it accommodates two helical pitch-lengths with $d \sim 2.55L$. Interestingly the $(\Delta\eta)_{Sm-N} / (\eta)_N \sim 2\%$ change of the packing fraction is a result of an expansion of the system by $(\Delta L_z)_{Sm-N} / (L_Z)_N \approx 7\%$ along the direction of the layer normal combined with a lateral contraction by $(\Delta L_\perp)_{Sm-N} / (L_\perp)_N \approx -2.4\%$. This inherently very anisotropic change of the volume of the system across the N-Sm transition explains the difficulty in obtaining a direct nematic to smectic phase transition during compression runs. The change of the packing fraction at the transition between the N phase at $P^* = 2.20$ and the isotropic phase at $P^* = 2.15$ is about $10\%$, a relatively high value that reflects the high compressibility of both phases close their phase transition. However, contrary to the considerably anisotropic expansion of the volume across the Sm-N phase change, the expansion of the system across the N-I transition shows a practically isotropic volume change, $(\Delta L_z)_{N-I} / (L_Z)_I \approx (\Delta L_\perp)_{N-I} / (L_\perp)_I$.

The graphs at the bottom of Fig. SI.2 show the evolution of the main global orientational order parameters that measure the average (over the entire simulation box) degree of the nematic like orientational order (see A.4) of the long molecular axis ($\overline{S_{\parallel}^{(z)}}$), the biaxiality of the phase ($\overline{\Delta S_{\perp}^{(z)}}$), and its form chirality ($\overline{S^*}$). Apparently, the Sm-N phase transition is a transformation from the high-pressure phase (smectic), which is biaxial and achiral, to a low-pressure phase (nematic) which exhibits form chirality and is globally uniaxial.

## S.2 Orientational Order Parameters for Curved Rod Nematics Showing Local $C_2$ Symmetry.

### S2.1 Order Parameters of First Rank.

There are generally 9 order parameters of first rank, $\langle \hat{a} \cdot \hat{A} \rangle$, with the angular brackets indicating ensemble averaging, $\hat{a}$ denoting any one of the three molecular axes, $\hat{x}, \hat{y}, \hat{z}$ and $\hat{A}$ any of the local frame axes $\hat{h}, \hat{l}, \hat{m}$, with $\hat{m}$ chosen to be the polar director and $\hat{h}$ the direction of the modulation (see also S3.3). As a result of the combination of the local $C_2$ symmetry of the phase and the $C_{2V}$ symmetry of the bent-rod molecules, the only non-vanishing of the 9 order parameters is $p_{\perp} \equiv \langle \hat{y} \cdot \hat{m} \rangle$.

### S2.2 Order Parameters of Second Rank.

The second rank parameters describing the orientational order of the three molecular axes $\hat{a} = \hat{x}, \hat{y}, \hat{z}$ are generally $S_{AB}^{(a)} = \left\langle 3\left(\hat{a} \cdot \hat{A}\right)\left(\hat{a} \cdot \hat{B}\right) - \delta_{A,B} \right\rangle / 2$. These form three sets (one for each molecular axis $\hat{a}$) of symmetric "ordering matrix" elements. Of those 18 elements in total, when expressed in the frame $\hat{h}, \hat{l}, \hat{m}$, the following 12 survive the implications ($S_{hm}^{(a)} = S_{lm}^{(a)} = 0$) of local phase symmetry: $S_{hh}^{(a)}, S_{ll}^{(a)}, S_{mm}^{(a)}, S_{hl}^{(a)}$ and of those, only 5 are independent, due to the geometrical identities

$$\sum_{a=x,y,z} S_{hh}^{(a)} = \sum_{a=x,y,z} \left(S_{mm}^{(a)} - S_{ll}^{(a)}\right) = \sum_{a=x,y,z} S_{hl}^{(a)} = S_{hh}^{(a)} + S_{ll}^{(a)} + S_{mm}^{(a)} = 0 \tag{S1}$$

Each of the above three ordering matrices, $S_{AB}^{(a)}$, can readily be diagonalized to obtain the principal axes and corresponding principal values : $\hat{m}$, being the local phase-symmetry axis, automatically constitutes a principal axis of all the second rank tensors; therefore the other two principal axes, say $\hat{h}_p^{(a)}, \hat{l}_p^{(a)}$ are obtained by a single rotation of $\hat{h}, \hat{l}$ about $\hat{m}$ by an angle $\theta^{(a)}$ satisfying the condition $S_{h_p^{(a)} l_p^{(a)}}^{(a)} = 0$. The value of this angle is determined from the values of the order parameters in the $\hat{h}, \hat{l}, \hat{m}$ frame according to

$$\tan 2\theta^{(a)} = -4S_{hl}^{(a)} \left[3S_{hh}^{(a)} + S_{mm}^{(a)} - S_{ll}^{(a)}\right]^{-1}. \tag{S2}$$

Obviously, the values of $S_{mm}^{(a)}$ remain unaffected by this rotation. The principal values corresponding to the axes $\hat{h}_p^{(a)}, \hat{l}_p^{(a)}$ are given by

$$
\begin{aligned}
S_{h_p h_p}^{(a)} &= \frac{1}{2}\left[ S_{hh}^{(a)}\left(1+\frac{1}{\cos 2\theta^{(\alpha)}}\right) + S_{ll}^{(a)}\left(1-\frac{1}{\cos 2\theta^{(\alpha)}}\right)\right] \\
S_{l_p l_p}^{(a)} &= \frac{1}{2}\left[ S_{ll}^{(a)}\left(1+\frac{1}{\cos 2\theta^{(\alpha)}}\right) + S_{hh}^{(a)}\left(1-\frac{1}{\cos 2\theta^{(\alpha)}}\right)\right]
\end{aligned} \quad . \tag{S3}
$$

The three principal values $S_{h_p h_p}^{(a)}, S_{l_p l_p}^{(a)}, S_{mm}^{(a)}$ are not independent as their sum vanishes. Accordingly, the ordering tensor in its principal axes frame is often represented by the largest of the three principal values, defining the ordering $S_{\|}^{(a)} \equiv \left\langle 3\left(\hat{e}_{\|}^{(a)} \cdot \hat{a}\right)^2 - 1 \right\rangle / 2$ along the major principal axis, $\hat{e}_{\|}^{(a)}$, and the difference of the other two, in ascending order, defining the biaxiality $\Delta S_{\perp}^{(a)} \equiv (3/2)\left\langle \left(\hat{e}_{\perp'}^{(a)} \cdot \hat{a}\right)^2 - \left(\hat{e}_{\perp''}^{(a)} \cdot \hat{a}\right)^2 \right\rangle$ with respect to the principal axes $\hat{e}_{\perp'}^{(a)}, \hat{e}_{\perp''}^{(a)}$. In this representation, the three principal axes $\hat{e}_{\|}^{(a)}$, $\hat{e}_{\perp'}^{(a)}, \hat{e}_{\perp''}^{(a)}$ simply constitute a relabelling of the axes $\hat{h}_p^{(a)}, \hat{l}_p^{(a)}, \hat{m}$ in ascending sequence of the respective principal values $S_{h_p h_p}^{(a)}, S_{l_p l_p}^{(a)}, S_{mm}^{(a)}$.

**S2.3 Vector-pseudovector order parameters.**

The order parameters $S_{AB}^{(a)}$ fully describe orientational ordering in nonpolar nonchiral LC phases to lowest (2nd) rank; the 1st rank $\langle a_A \rangle$ describe additional features of the polar phases. Order parameters detecting the existence of structural chirality in the phase can be formulated. Thus, the local $C_2$ phase symmetry combined with the $C_{2V}$ molecular symmetry, give rise to nontrivial values for some components of the vector-pseudovector ordering tensors $S_{AB}^{*(a)}$. With the *y*-axis assigned as the $C_2$ molecular axis, these are defined as

$$
S_{AB}^{*(a)} \equiv \frac{1}{2}\left\langle \left(\hat{a}\cdot\hat{A}\right)\left[\left(\hat{a}\times\hat{y}\right)\cdot\hat{B}\right] + \left(\hat{a}\cdot\hat{B}\right)\left[\left(\hat{a}\times\hat{y}\right)\cdot\hat{A}\right]\right\rangle, \tag{S4}
$$

they are represented by symmetric and traceless matrices and present invariance under improper rotations of the molecule. Clearly $S_{AB}^{*(y)}$ is null and therefore $S_{AB}^{*(z)} = -S_{AB}^{*(x)} \equiv S_{AB}^{*}$. By the $C_2$ local symmetry of the phase, the only non-vanishing components of the vector-pseudovector ordering tensor in the $\hat{h},\hat{l},\hat{m}$ frame are $S^*_{hh},S^*_{ll},S^*_{mm},S^*_{hl}$ . Due to the identity $S^*_{hh}+S^*_{ll}+S^*_{mm}=0$, there are only three independent vector-pseudovector order parameters which can be conveniently chosen to be $S^*_{hh};S^*_{mm}-S^*_{ll};S^*_{hl}$. The first two of these components describe correlations between the tilt of the molecular axes $\hat{x},\ \hat{z}$ and the polar order of the molecular symmetry axis $\hat{y}$ in the local $\hat{h},\hat{l},\hat{m}$ phase axes, while the third includes, in addition, polarity - biaxiality correlations.

Consider $S^*_{hh}$. With $\vec{t}_h^{(z)}=\left(\hat{h}\cdot\hat{z}\right)\left(\hat{h}\times\hat{z}\right)$ denoting the pseudovector describing the angular deviation (instantaneous "tilt") of the molecular axis $\hat{z}$ relative to the phase axis $\hat{h}$ it follows that

$$S^*_{hh}=\left\langle\vec{t}_h^{(z)}\cdot\hat{y}\right\rangle=\left\langle\left(\vec{t}_h^{(z)}\cdot\hat{m}\right)\left(\hat{y}\cdot\hat{m}\right)\right\rangle+\left\langle\left(\vec{t}_h^{(z)}\cdot\hat{l}\right)\left(\hat{y}\cdot\hat{l}\right)\right\rangle. \tag{S5}$$

Accordingly, $S^*_{hh}$ measures the ensemble average of the projection of the molecular symmetry axis $\hat{y}$ along the tilt pseudovector $\vec{t}_h^{(z)}$, or , equivalently, the correlations between the tilt components along the axes normal to $\hat{h}$ (i.e. $\hat{l}$ and $\hat{m}$) and the respective polar ordering of the molecular symmetry axis along those phase axes. Note that while local phase symmetry implies $\left\langle\left(\vec{t}_h^{z}\cdot\hat{l}\right)\right\rangle=0=\left\langle\left(\hat{y}\cdot\hat{l}\right)\right\rangle$, there is no such implication for either of $\left\langle\left(\vec{t}_h^{(z)}\cdot\hat{m}\right)\right\rangle$ or $\left\langle\left(\hat{y}\cdot\hat{m}\right)\right\rangle$.

Analogously for the order parameters $S^*_{ll},S^*_{mm}$ we obtain

$$S^*_{mm}=\left\langle\vec{t}_m^{(z)}\cdot\hat{y}\right\rangle=\left\langle\left(\vec{t}_m^{(z)}\cdot\hat{h}\right)\left(\hat{y}\cdot\hat{h}\right)\right\rangle+\left\langle\left(\vec{t}_m^{(z)}\cdot\hat{l}\right)\left(\hat{y}\cdot\hat{l}\right)\right\rangle \tag{S6}$$

and

$$S^*_{ll}=\left\langle\vec{t}_l^{(z)}\cdot\hat{y}\right\rangle=\left\langle\left(\vec{t}_l^{(z)}\cdot\hat{h}\right)\left(\hat{y}\cdot\hat{h}\right)\right\rangle+\left\langle\left(\vec{t}_l^{(z)}\cdot\hat{m}\right)\left(\hat{y}\cdot\hat{m}\right)\right\rangle, \tag{S7}$$

showing the respective tilt-polarity correlations relative to the appropriate phase axes.

The $S^*_{hl}$ order parameter describes additional correlations,

$$\begin{aligned}S^*_{lh}&=\frac{1}{2}\left\langle\left(\hat{z}\cdot\hat{h}\right)\left[\left(\hat{z}\times\hat{y}\right)\cdot\hat{l}\right]+\left(\hat{z}\cdot\hat{l}\right)\left[\left(\hat{z}\times\hat{y}\right)\cdot\hat{h}\right]\right\rangle\\&=-\frac{1}{2}\left\langle\left(\hat{y}\cdot\hat{h}\right)\left(\vec{t}_h^{z}\cdot\hat{l}\right)+\left(\hat{y}\cdot\hat{l}\right)\left(\vec{t}_l^{z}\cdot\hat{h}\right)+\left(\hat{y}\cdot\hat{m}\right)\left[\left(\hat{z}\cdot\hat{h}\right)^2-\left(\hat{z}\cdot\hat{l}\right)^2\right]\right\rangle\end{aligned} \tag{S8}$$

that is, polarity-biaxiality correlations, in addition to tilt-polarity correlations referring to crossed

($\hat{l},\hat{h}$ and $\hat{h},\hat{l}$) axes.

The matrix associated with the ordering tensor can be diagonalized in analogy to Eqs(S2 and 3) through a rotation about $\hat{m}$ to obtain the principal axes $\hat{h}_p^*,\hat{l}_p^*$, the rotation angle $\theta^*$ and the respective principal values with $S^*_{h_p h_p}+S^*_{l_p l_p}+S^*_{mm}=0$.

## S.3 Orientational Pair Correlation Functions for Curved Rod Nematics of Local $C_2$ Symmetry and Roto-Translational Modulations in One Dimension

The calculation of local order from simulations of periodically modulated phases encounters certain technical ambiguities stemming from the interplay between the periodicity of the modulation and the periodic boundary conditions imposed on the simulated system. Pair correlation functions offer a consistent way to remove such ambiguities from the calculation.

In the case of the one-dimensional modulations of the present systems this can be readily understood: With the modulation taking place along the $\hat{Z}$ -axis, the order parameters $\langle a_A \rangle$, $S^{(a)}_{AB}$, and $S^*_{AB}$ are formed by ensemble averages of various combinations of the projections of the molecular axes *a* along phase axes *A,B*, at some slab of fixed *Z*-value. However, the respective correlation functions involve projections of the molecular axes of molecules in a given slab at $Z_1$ along the molecular axes of other molecules located in a second slab at $Z_2$ and thus depends on the separation $Z_1$-$Z_2$, irrespective of the absolute positions of the two slabs.

The pair correlation function associated with the polar order parameter $p_\perp \equiv \langle \hat{y}\cdot\hat{m}\rangle$ are presented in the main text. The functions associated with $S^{(a)}_{AB}$, and $S^*_{AB}$ are defined as follows.

### S3.1 Functions describing the correlation of 2nd rank ordering.

These are defined by:

$$g_2^{(ac;bd)}(Z)=\frac{\left\langle \sum_{i,j}\left(\frac{3}{2}\left(\hat{a}(\mathbf{r}_i)\cdot\hat{b}(\mathbf{r}_j)\right)\left(\hat{c}(\mathbf{r}_i)\cdot\hat{d}(\mathbf{r}_j)\right)-\frac{1}{2}\delta_{ac}\delta_{bd}\right)\delta(\mathrm{Z}-\mathbf{r}_{ij}\cdot\hat{Z})\right\rangle}{\left\langle \sum_{i,j}\delta(\mathrm{Z}-\mathbf{r}_{ij}\cdot\hat{Z})\right\rangle}, \quad \text{(S9)}$$

where $\hat{a},\hat{b},\hat{c},\hat{d}$ denote molecular axis directions at the indicated positions. For molecules of $C_{2V}$ symmetry, such as the bent tubes in Figure 2 of the main text, the surviving functions are those with *a*=*c* and *b*=*d* and therefore there are 9 non-vanishing correlation functions of rank 2. Moreover, as a result of the 6 identities

$$\sum_a g_2^{(aa;bb)}(Z) = 0 = \sum_b g_2^{(aa;bb)}(Z); \quad a,b = x,y,z \,, \qquad , \tag{S10}$$

only three of these non-vanishing functions are independent.

**S3.2 Functions describing the correlation of vector-pseudovector ordering.**

With $\hat{y}$ chosen to denote the molecular symmetry axis, these are generally defined by

$$g^{*(ac;bd)}(Z) = \frac{3}{2}\frac{\left\langle \sum_{i,j}\begin{pmatrix}\left(\hat{a}(\mathbf{r}_i)\cdot\hat{b}(\mathbf{r}_j)\right)\left(\left(\hat{c}(\mathbf{r}_i)\times\hat{y}(\mathbf{r}_i)\right)\cdot\left(\hat{d}(\mathbf{r}_j)\times\hat{y}(\mathbf{r}_j)\right)\right)+\\ \left(\hat{a}(\mathbf{r}_i)\cdot\hat{d}(\mathbf{r}_j)\right)\left(\left(\hat{c}(\mathbf{r}_i)\times\hat{y}(\mathbf{r}_i)\right)\cdot\left(\hat{b}(\mathbf{r}_j)\times\hat{y}(\mathbf{r}_j)\right)\right)\end{pmatrix}\delta(\mathrm{Z}-\mathbf{r}_{ij}\cdot\hat{Z})\right\rangle}{\left\langle \sum_{i,j}\delta(\mathrm{Z}-\mathbf{r}_{ij}\cdot\hat{Z})\right\rangle}\,, \tag{S11}$$

in the notation of eqs(S9 and 10). Here only the *a*=*c* and *b*=*d* combinations survive the $C_2$ molecular symmetry. Clearly $g^{*(yy;aa)}(Z) = g^{*(aa;yy)}(Z) = 0$.

Furthermore, $g^{*(zz;zz)}(Z) = g^{*(xx;xx)}(Z)$ and the identity

$$\sum_a g^{*(aa;bb)}(Z) = \sum_b g^{*(aa;bb)}(Z) = \frac{2}{3} g_1^{(yy)}(Z)\,, \tag{S12}$$

imply that, given $g_1^{(yy)}(Z)$, there is essentially just one independent vector-pseudovector correlation function, which, for simplicity, is denoted by $g^*(Z) \equiv g^{*(zz;zz)}(Z)$.

**S3.3 The roto-translation model of modulated order.**

Briefly, this one-dimensional modulation model, predicted by the molecular theory of the polar-twisted nematic phase ($N_{PT}$) [40], is formulated by assuming that at each Z position (along the modulation axis $\hat{Z}$) there is a local frame of axes with respect to which the principal values of the order parameters are the same, independent of Z . What changes on translating along Z is the direction of the respective principal axes, following a simple rotation about $\hat{Z}$ by an angle $\varphi$, proportional to the translation, i.e. $\varphi = kZ$, with $k \equiv 2\pi / d$ the wavenumber of the roto-translation modulation and $d$ its periodicity length ("pitch").

With $\hat{m}(Z)$ defining the loacal direction of maximal alignment, and choosing the $\hat{h}$ of the local frame to be parallel to the modulation direction $\hat{Z}$, in accord with $\hat{m}(Z)$ remaining perpendicular to the modulation direction, the roto-translational modulation implies that, on moving along the modulation axis, (i) the two local axes $\hat{m},\hat{l}$ rotate about that direction by an angle proportional to the displacement of the modulation, i.e.

$$\hat{m}(Z) = \hat{Y}\cos(kZ) - \hat{X}\sin(kZ) \; ; \; \hat{l}(Z) = \hat{X}\cos(kZ) + \hat{Y}\sin(kZ), \quad \text{(S13)}$$

and (ii) the values of the order parameters measuring the degree of orientational ordering of the molecules along the axes $\hat{h},\hat{l},\hat{m}$ remain invariant to displacements along $\hat{Z} \parallel \hat{h}$. The choice of the $\hat{X},\hat{Y}$ axes, to form the orthogonal macroscopic frame $\hat{X},\hat{Y},\hat{Z}$, is obviously coupled to the choice of the origin $Z = 0$.

**S3.4 The form of the correlation functions according to the roto-translation model.**

The detailed information obtained directly from the simulations for the correlation functions, regarding the structure of the modulated nematic and the local order parameters of the phase, is used to test the predictions of the roto-translation model. Perfect agreement is found for all the correlation functions computed.

Consider, for example, $g_1^{(yy)}(Z)$. It can be calculated theoretically assuming roto-translational symmetry of the polar axis $\hat{m}$ along $\hat{h}$. Expressing the molecular axes $\hat{y}_i, \hat{y}_j$ in the local axes associated with two planes perpendicular to $\hat{h}$ and positioned at $Z = Z_1$ and $Z = Z_2$ we have $\hat{y}_i \cdot \hat{y}_j = \left[ (\hat{y}_i \cdot \hat{m}(Z_1))(\hat{y}_j \cdot \hat{m}(Z_2)) + (\hat{y}_i \cdot \hat{l}(Z_1))(\hat{y}_j \cdot \hat{l}(Z_2)) \right] \cos k(Z_2 - Z_1) + (\hat{y}_i \cdot \hat{h})(\hat{y}_j \cdot \hat{h})$.

Taking then into account the $C_2$ symmetry about $\hat{m}$ we obtain upon averaging

$\left\langle \sum_{i,j} \hat{y}_i \cdot \hat{y}_j \right\rangle / \left\langle \sum_{i,j} \right\rangle = \langle \hat{y} \cdot \hat{m} \rangle^2 \cos k(Z_2 - Z_1)$. Therefore, with the notation introduced in S2.1 for the polar order parameter we have

$$g_1^{(yy)}(Z) = p_\perp^2 \cos kZ \qquad . \tag{S14}$$

Theoretical expressions, within the roto-translation model, for the three independent correlation functions $g_2^{(xx)}(Z)$, $g_2^{(yy)}(Z)$ and $g_2^{(zz)}(Z)$ can be obtained by an analogous procedure. Consider for example $g_2^{(xx)}(Z)$. On expressing the orientations of $\hat{x}_i, \hat{x}_j$ in local axes associated with planes at $Z = Z_1$ and $Z = Z_2$ we have, taking into account the local C2 symmetry about $\hat{m}$,

$$\left\langle \left[ \sum_{i,j} \left( \frac{3}{2} \left( \hat{x}_i \cdot \hat{x}_j \right)^2 - \frac{1}{2} \right) \right] / \left( \sum_{i,j} \right) \right\rangle =$$
$$= \left\langle \frac{3}{2} \left( \hat{x} \cdot \hat{h} \right)^2 - \frac{1}{2} \right\rangle^2 + 3 \left\langle \left( \hat{x} \cdot \hat{h} \right) \left( \hat{x} \cdot \hat{l} \right) \right\rangle^2 \cos k \left( Z_2 - Z_1 \right) + \frac{3}{4} \left( \left\langle \left( \hat{x} \cdot \hat{m} \right)^2 \right\rangle - \left\langle \left( \hat{x} \cdot \hat{l} \right)^2 \right\rangle \right)^2 \cos 2k \left( Z_2 - Z_1 \right)$$

and similarly for the $\hat{y}, \hat{z}$ axes. Therefore, we have the final expressions

$$g_2^{(aa)}(Z) = \left( S_{hh}^{(a)} \right)^2 + \frac{4}{3} \left( S_{lh}^{(a)} \right)^2 \cos kZ + \frac{1}{3} \left( S_{mm}^{(a)} - S_{ll}^{(a)} \right)^2 \cos 2kZ, \; for \; a = x, y, z \; . \tag{S15}$$

This relates the correlation functions to the values of all the second rank orientational order parameters that describe the C2 local orientational order, and verifies that the order parameter values are indeed consistent with the constraints in Eq(S1). Note that opposite signs of the order parameter $S_{lh}^{(a)}$ correspond to domains of opposite chirality.

Similarly, the vector-pseudovector correlation function $g^*(Z)$ has the following theoretical expression within the roto-translation model

$$g^*(Z) = \left( S_{hh}^* \right)^2 + \left[ \frac{4}{3} \left( S_{hl}^* \right)^2 + \frac{3}{4} \left\langle \hat{y} \cdot \hat{m} \right\rangle^2 \right] \cos kZ + \frac{1}{3} \left( S_{mm}^* - S_{ll}^* \right)^2 \cos 2kZ \qquad . \tag{S16}$$

## S.4 Modulated and Global Orientational Order

In view of the eqs(S13), describing the transformation of the local axes implied by the roto-translation symmetry, the order parameters of section S.2 can be expressed in the phase-fixed macroscopic frame $\hat{X}, \hat{Y}, \hat{Z}$ as follows:

$$p_X = p_\perp \sin(kZ); \; p_Y = p_\perp \cos(kZ); \; p_Z = 0, \tag{S17}$$

for the polar order parameter in S2.1.

$$S_{AB}^{(a)}(Z)=\begin{pmatrix} -\frac{1}{2}S_{hh}^{(a)}+\frac{1}{2}\left(S_{ll}^{(a)}-S_{mm}^{(a)}\right)\cos 2kZ & \frac{1}{2}\left(S_{ll}^{(a)}-S_{mm}^{(a)}\right)\sin 2kZ & S_{lh}^{(a)}\cos kZ \\ \frac{1}{2}\left(S_{ll}^{(a)}-S_{mm}^{(a)}\right)\sin 2kZ & -\frac{1}{2}S_{hh}^{(a)}-\frac{1}{2}\left(S_{ll}^{(a)}-S_{mm}^{(a)}\right)\cos 2kZ & S_{lh}^{(a)}\sin kZ \\ S_{lh}^{(a)}\cos kZ & S_{lh}^{(a)}\sin kZ & S_{hh}^{(a)} \end{pmatrix} \quad \text{(S18)}$$

for the 2nd rank ordering tensors in section S2.2, with $A,B=X,Y,Z$, and similarly for the vector-paseudovector order paramaters in section S2.3, only with the superrscript $(a)=(x),(y),(z)$ replaced by *.

On averaging Eq(S17) over the Z-variable within a full pitch length $d=2\pi/k$, both alternating components of the polar order parameter yield vanishing average values, i.e. $\overline{p_X}=\overline{p_Y}=0$, with the bar indicating Z-averaging. The same averaging of the expression in Eq (S18) yields the uniaxial ordering tensors

$$\overline{S_{AB}^{(a)}}=S_{hh}^{(a)}\begin{pmatrix} -\frac{1}{2} & 0 & 0 \\ 0 & -\frac{1}{2} & 0 \\ 0 & 0 & 1 \end{pmatrix}; \quad \overline{S_{AB}^{*}}=S_{hh}^{*}\begin{pmatrix} -\frac{1}{2} & 0 & 0 \\ 0 & -\frac{1}{2} & 0 \\ 0 & 0 & 1 \end{pmatrix}, \quad \text{(S19)}$$

with principal axis along $\hat{Z}$.

The components of the Z-averaged tensors can be independently calculated from the simulations by sampling within the entire simulation box. For example, the average ordering tensor $\overline{S_{AB}^{(a)}}$ of the MC generated configurations of the systems is calculated as the average over MC cycles of $\frac{1}{N}\sum_{i=1}^{N}\left(3a_{i,A}a_{i,B}-\delta_{A,B}\right)/2$, where $N$ is the number of particles and $i$ runs over all the particles.

The eigenvalues, and the corresponding eigenvectors, of the resulting $\overline{S_{AB}^{(a)}}$ are calculated and sorted out in descending order.

Direct spatial averaging of the orientational correlation functions yields, according to Eqs(S15 and 16) $\overline{g_2^{(aa)}} = \left(S_{hh}^{(a)}\right)^2$ and $\overline{g_2^*} = \left(S_{hh}^*\right)^2$. These relations are reproduced consistently with those in Eq (S19) by direct calculation from the simulations at various pressures.